\documentclass[reprint,superscriptaddress,amsmath,amssymb,aps,longbibliography,floatfix]{revtex4-1}

\usepackage[colorlinks]{hyperref}
\usepackage{graphicx}
\usepackage{dcolumn}
\usepackage{bm}
\usepackage{booktabs}
\usepackage{array}
\usepackage{color}
\usepackage{bm}
\usepackage{verbatim}
\usepackage{cleveref}
\usepackage{lipsum}
\usepackage{algorithm}
\usepackage{algpseudocode}

\newcommand{\eq}[1]{Eq.~\hyperref[eq:#1]{(\ref*{eq:#1})}}
\renewcommand{\sec}[1]{\hyperref[sec:#1]{Section~\ref*{sec:#1}}}
\newcommand{\app}[1]{\hyperref[app:#1]{Appendix~\ref*{app:#1}}}
\newcommand{\tab}[1]{\hyperref[tab:#1]{Table~\ref*{tab:#1}}}
\newcommand{\fig}[1]{\hyperref[fig:#1]{Figure~\ref*{fig:#1}}}
\newcommand{\figa}[2]{\hyperref[fig:#1]{Figure~\ref*{fig:#1}#2}}
\newcommand{\figx}[2]{\hyperref[fig:#1]{Figure~\ref*{fig:#1}(#2)}}
\newcommand{\thm}[1]{\hyperref[thm:#1]{Theorem~\ref*{thm:#1}}}
\newcommand{\lem}[1]{\hyperref[lem:#1]{Lemma~\ref*{lem:#1}}}
\newcommand{\cor}[1]{\hyperref[cor:#1]{Corollary~\ref*{cor:#1}}}
\newcommand{\defn}[1]{\hyperref[def:#1]{Definition~\ref*{def:#1}}}
\newcommand{\alg}[1]{\hyperref[alg:#1]{Algorithm~\ref*{alg:#1}}}

\newcommand{\syc}{\textsc{syc}}
\newcommand{\swap}{\textsc{swap}}
\newcommand{\iswap}{\rm{i}\textsc{swap}}
\newcommand{\sqrtiswap}{\sqrt{\iswap}}
\newcommand{\cphase}{\textsc{cphase}}
\newcommand{\zzswap}{e^{-i \gamma w ZZ}\! \cdot \swap}
\newcommand{\PhX}{\textrm{PhX}}

\newcommand{\gammavector}{\boldsymbol{\gamma}}
\newcommand{\betavector}{\boldsymbol{\beta}}
\newcommand{\bra}[1]{\langle #1|}
\newcommand{\ket}[1]{|#1\rangle}
\newcommand{\cmin}{C_\mathrm{min}}
\newcommand{\C}{\langle C \rangle}
\newcommand{\covercmin}{\C / \cmin}

\begin{document}

\title{Quantum Approximate Optimization of Non-Planar Graph Problems\\ on a Planar Superconducting Processor}

\newcommand{\Google}{\affiliation{Google Quantum AI}}
\newcommand{\UMass}{\affiliation{Department of Electrical and Computer Engineering, University of Massachusetts, Amherst, MA}}
\newcommand{\UCSB}{\affiliation{Department of Physics, University of California, Santa Barbara, CA}}
\newcommand{\UCR}{\affiliation{Department of Electrical and Computer Engineering, University of California, Riverside, CA}}
\newcommand{\VW}{\affiliation{Volkswagen Data:Lab}}
\newcommand{\NASA}{\affiliation{NASA Ames Research Center, Moffett Field, CA}}
\newcommand{\UCB}{\affiliation{Department of Electrical Engineering and Computer Sciences, University of California, Berkeley, CA}}
\newcommand{\FAU}{\affiliation{Friedrich-Alexander University Erlangen-Nürnberg, Department of Physics, Erlangen, Germany}}
\newcommand{\UM}{\affiliation{Department of Electrical Engineering and Computer Science, University of Michigan, Ann Arbor, MI}}
\newcommand{\Harvard}{\affiliation{Department of Physics, Harvard University, Cambridge, MA}}
\newcommand{\Leiden}{\affiliation{Leiden University, Leiden, Netherlands}}


\author{Matthew P.~Harrigan}
\email[Corresponding author (M.~Harrigan): ]{mpharrigan@google.com}
\Google

\author{Kevin J.~Sung}
\Google
\UM

\author{Matthew Neeley}
\Google

\author{Kevin J.~Satzinger}
\Google

\author{Frank Arute}
\Google

\author{Kunal Arya}
\Google

\author{Juan Atalaya}
\Google

\author{Joseph C.~Bardin}
\Google
\UMass

\author{Rami Barends}
\Google

\author{Sergio Boixo}
\Google

\author{Michael Broughton}
\Google

\author{Bob B.~Buckley}
\Google

\author{David A.~Buell}
\Google

\author{Brian Burkett}
\Google

\author{Nicholas Bushnell}
\Google

\author{Yu Chen}
\Google

\author{Zijun Chen}
\Google

\author{Ben Chiaro}
\Google
\UCSB

\author{Roberto Collins}
\Google

\author{William Courtney}
\Google

\author{Sean Demura}
\Google

\author{Andrew Dunsworth}
\Google

\author{Daniel Eppens}
\Google

\author{Austin Fowler}
\Google

\author{Brooks Foxen}
\Google

\author{Craig Gidney}
\Google

\author{Marissa Giustina}
\Google

\author{Rob Graff}
\Google

\author{Steve Habegger}
\Google

\author{Alan Ho}
\Google

\author{Sabrina Hong}
\Google

\author{Trent Huang}
\Google

\author{L. B.~Ioffe}
\Google

\author{Sergei V.~Isakov}
\Google

\author{Evan Jeffrey}
\Google

\author{Zhang Jiang}
\Google

\author{Cody Jones}
\Google

\author{Dvir Kafri}
\Google

\author{Kostyantyn Kechedzhi}
\Google

\author{Julian Kelly}
\Google

\author{Seon Kim}
\Google

\author{Paul V.~Klimov}
\Google

\author{Alexander N.~Korotkov}
\Google
\UCR

\author{Fedor Kostritsa}
\Google

\author{David Landhuis}
\Google

\author{Pavel Laptev}
\Google

\author{Mike Lindmark}
\Google

\author{Martin Leib}
\VW

\author{Orion Martin}
\Google

\author{John M.~Martinis}
\Google
\UCSB

\author{Jarrod R.~McClean}
\Google

\author{Matt McEwen}
\Google
\UCSB

\author{Anthony Megrant}
\Google

\author{Xiao Mi}
\Google

\author{Masoud Mohseni}
\Google

\author{Wojciech Mruczkiewicz}
\Google

\author{Josh Mutus}
\Google

\author{Ofer Naaman}
\Google

\author{Charles Neill}
\Google

\author{Florian Neukart}
\VW

\author{Murphy Yuezhen Niu}
\Google

\author{Thomas E.~O'Brien}
\Google

\author{Bryan O'Gorman}
\NASA
\UCB

\author{Eric Ostby}
\Google

\author{Andre Petukhov}
\Google

\author{Harald Putterman}
\Google

\author{Chris Quintana}
\Google

\author{Pedram Roushan}
\Google

\author{Nicholas C.~Rubin}
\Google

\author{Daniel Sank}
\Google

\author{Andrea Skolik}
\VW
\Leiden

\author{Vadim Smelyanskiy}
\Google

\author{Doug Strain}
\Google

\author{Michael Streif}
\VW
\FAU

\author{Marco Szalay}
\Google

\author{Amit Vainsencher}
\Google

\author{Theodore White}
\Google

\author{Z.~Jamie Yao}
\Google

\author{Ping Yeh}
\Google

\author{Adam Zalcman}
\Google

\author{Leo Zhou}
\Google
\Harvard

\author{Hartmut Neven}
\Google

\author{Dave Bacon}
\Google

\author{Erik Lucero}
\Google

\author{Edward Farhi}
\Google

\author{Ryan Babbush}
\email[Corresponding author (R.~Babbush): ]{babbush@google.com}
\Google

\date{\today}

\begin{abstract}
We demonstrate the application of the Google Sycamore superconducting qubit quantum processor to combinatorial optimization problems with the quantum approximate optimization algorithm (QAOA). Like past QAOA experiments, we study performance for problems defined on the (planar) connectivity graph of our hardware; however, we also apply the QAOA to the Sherrington-Kirkpatrick model and MaxCut, both high dimensional graph problems for which the QAOA requires significant compilation. Experimental scans of the QAOA energy landscape show good agreement with theory across even the largest instances studied (23 qubits) and we are able to perform variational optimization successfully. For problems defined on our hardware graph we obtain an approximation ratio that is independent of problem size and observe, for the first time, that performance increases with circuit depth.  For problems requiring compilation, performance decreases with problem size but still provides an advantage over random guessing for circuits involving several thousand gates. This behavior highlights the challenge of using near-term quantum computers to optimize problems on graphs differing from hardware connectivity. As these graphs are more representative of real world instances, our results advocate for more emphasis on such problems in the developing tradition of using the QAOA as a holistic, device-level benchmark of quantum processors.
\end{abstract}

\maketitle

\section{Introduction}

The Google Sycamore superconducting qubit platform has been used to demonstrate computational capabilities surpassing those of classical supercomputers for certain sampling tasks \cite{google_supremacy_2019}. However, it remains to be seen whether such processors will be able to achieve a similar computational advantage for problems of practical interest. Along with quantum chemistry \cite{aspuru_guzik_simulated_2005,omalley_scalable_2016}, machine learning \cite{2017-biamonte-quantum-machine-learning}, and simulation of physical systems \cite{lloyd_universal_1996}, discrete optimization has been widely anticipated as a promising area of application for quantum computers.

Beginning with a focus on quantum annealing \cite{kadowaki_annealing_1998} and adiabatic quantum computing \cite{farhi_applied_2001}, the possibility of quantum enhanced optimization has driven much interest in quantum technologies over the years. This is because faster optimization could prove transformative for diverse areas such as logistics, finance, machine learning, and more. Such discrete optimization problems can be expressed as the minimization of a quadratic function of binary variables \cite{lucas-ising,barahona_ising_1982}, and one can visualize these cost functions as graphs with binary variables as nodes and (weighted) edges connecting bits whose (weighted) products sum to the total cost function value. For most industrially-relevant problems, these graphs are non-planar and many ancilla would be required to embed them in (quasi-)planar graphs matching the qubit connectivity of most hardware platforms \cite{choi_minor_2008}. This limits the applicability of scalable architectures for quantum annealing \cite{denchev_tunneling_2016} and corresponds to increased circuit complexity in digital quantum algorithms for optimization such as QAOA.

\begin{figure*}[htbp]
    \centering
    \includegraphics[width=0.99\textwidth]{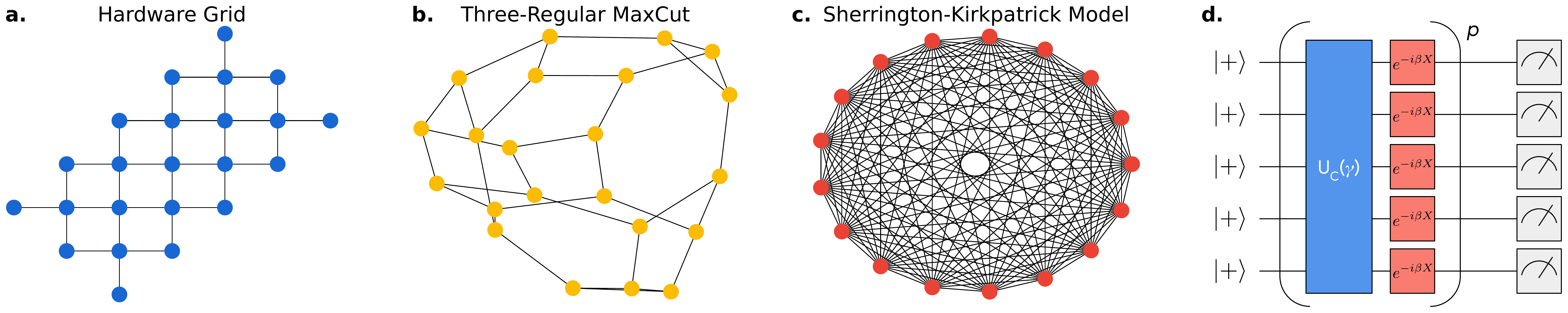}
    \caption{
    We studied three families of optimization problems: \textbf{a.} Hardware Grid problems with a graph matching the hardware connectivity of the 23 qubits used in this experiment. \textbf{b.} MaxCut on random 3-regular graphs, with the largest instance depicted (22 qubits). \textbf{c.} The fully-connected Sherrington-Kirkpatrick (SK) model shown at the largest size (17 qubits). \textbf{d.} QAOA uses $p$ applications of problem and driver unitaries to approximate solutions to optimization problems. The parameters $\gamma$ and $\beta$ are shared among qubits in a layer but different for each of the $p$ layers.
    }
    \label{fig:problems}
\end{figure*}

The quantum approximate optimization algorithm (QAOA) is the most studied gate model approach for optimization using near-term devices \cite{qaoa}. While the prospects for achieving quantum advantage with QAOA remain unclear, QAOA prescribes a simple paradigm for optimization which makes it amenable to both analytical results and implementation on current processors \cite{qaoa2, nasa-qc-perspective, wecker_training_2016, qaoa-supremacy, qaoa-grovers, qaoa-fermionic, alternating-operator-ansatz, qaoa-universal, farhi-sk-qaoa}. For these reasons, QAOA has also become popular as a system-level benchmark of quantum hardware. This work builds on prior experimental demonstrations of QAOA on superconducting qubits \cite{rigetti-qaoa-clustering, 2019-julich-qaoa-on-ibm, rigetti-xy-gate, chalmers-2q-qaoa},
ion traps \cite{2019-monroe-qaoa},
and photonics systems \cite{2018-bristol-photonics}.
We compare results from these past experiments in \app{compare}.

We are able to experimentally resolve, for the first time, increased performance with greater QAOA depth and apply QAOA to cost functions on graphs that deviate significantly from our hardware connectivity.
Owing to the low error rates of the Sycamore platform, the trade-off between the theoretical increase in quality of solutions with increasing QAOA depth and additional noise is apparent for hardware-native problems. We also apply the algorithm to non-native graph problems with their necessary compilation overhead and study the scaling of solution quality and problem size.
Our results reveal that the performance of QAOA is qualitatively different when applied to hardware native graphs versus more complex graphs, highlighting the challenge of scaling QAOA to problems of industrial importance.

For this study, we used a ``Sycamore'' quantum processor which consists of a two-dimensional array of 54 transmon qubits \cite{google_supremacy_2019}. Each qubit is tunably coupled to four nearest neighbors in a rectangular lattice. In this case, all device calibration was fully automated and data was collected using a cloud interface to the platform programmed using Cirq \cite{cirq}. Our experiment was restricted to 23 physical qubits of the larger Sycamore device, arranged in a topology depicted in \figa{problems}{a}.

The combinatorial optimization problems we study in this work are defined through a cost function $C(\mathbf{z})$ with a corresponding quantum operator $C$ given by
\begin{align}\label{eq:cost-ham}
    C  =  \sum_{j < k}  w_{jk}  Z_j  Z_k
\end{align}
where $\mathbf{z}$ is a classical bitstring, $Z_j$ denotes the Pauli $Z$ operator on qubit $j$ and the $w_{jk}$ correspond to scalar weights with values $\{0, \pm 1\}$. Because these clauses act on at most two qubits we are able to associate a graph with a given problem instance; if $w_{jk} \neq 0$, there is an edge between $j$ and $k$ in the graph. We will study three families of problem graphs depicted in \fig{problems}.

The shallowest depth version of QAOA consists of the application of two unitary operators. At higher depths the same two unitaries are sequentially reapplied but with different parameters. We denote the number of repeated application of this pair of unitaries as $p$, giving $2p$ parameters. The first unitary prescribed by QAOA is
\begin{align}\label{eq:problem-unitary}
    U_C \! \left(\gamma \right) = e^{-i \gamma C } = \prod_{j < k} e^{-i \gamma w_{jk} Z_j Z_k},
\end{align}
which depends on the parameter $\gamma$ and applies a phase to pairs of bits according to the problem-specific cost function.
The second operation is the driver unitary
\begin{align}\label{eq:driver-unitary}
    U_B \! \left(\beta \right) = e^{-i \beta B} = \prod_j e^{- i \beta X_j},
    \quad \qquad
    B = \sum_j X_j
\end{align}
where $X_j$ denotes the Pauli $X$ operator acting on qubit $j$. This unitary drives transitions between bitstrings within the superposition state. These operators can be implemented by sequentially evolving under each term of the cost function, as suggested by \eq{problem-unitary} and \eq{driver-unitary}.

For depth $p$ and $n$ qubits we prepare the state parameterized by $\gammavector = (\gamma_1, \dots, \gamma_p)$ and $\betavector = (\beta_1, \dots, \beta_p)$
\begin{align}\label{eq:gamma-beta-state}
	| \gammavector , \betavector \rangle = U_B(\beta_p)  U_C(\gamma_p ) \cdots U_B(\beta_1) U_C(\gamma_1 )  \ket{+}^{\otimes n},
\end{align}
where $\ket{+}^{\otimes n}$ is the symmetric superposition of all $2^n$ computational basis states. The application of the QAOA circuit to this initial state is depicted in \figa{problems}{d}. For a given $p$, we can find parameters to minimize the expectation value of the cost
\begin{equation}
\C = \langle \gammavector , \betavector | C | \gammavector, \betavector \rangle.
\end{equation} For comparison among problem instances, we divide by $\cmin = \min_\mathbf{z} C(\mathbf{z})$, which is negative for all problems we study, so we are in fact maximizing $\covercmin$.

\section{Compilation and Problem Families}

\begin{figure}[htbp]
    \centering
    \includegraphics[width=0.48\textwidth]{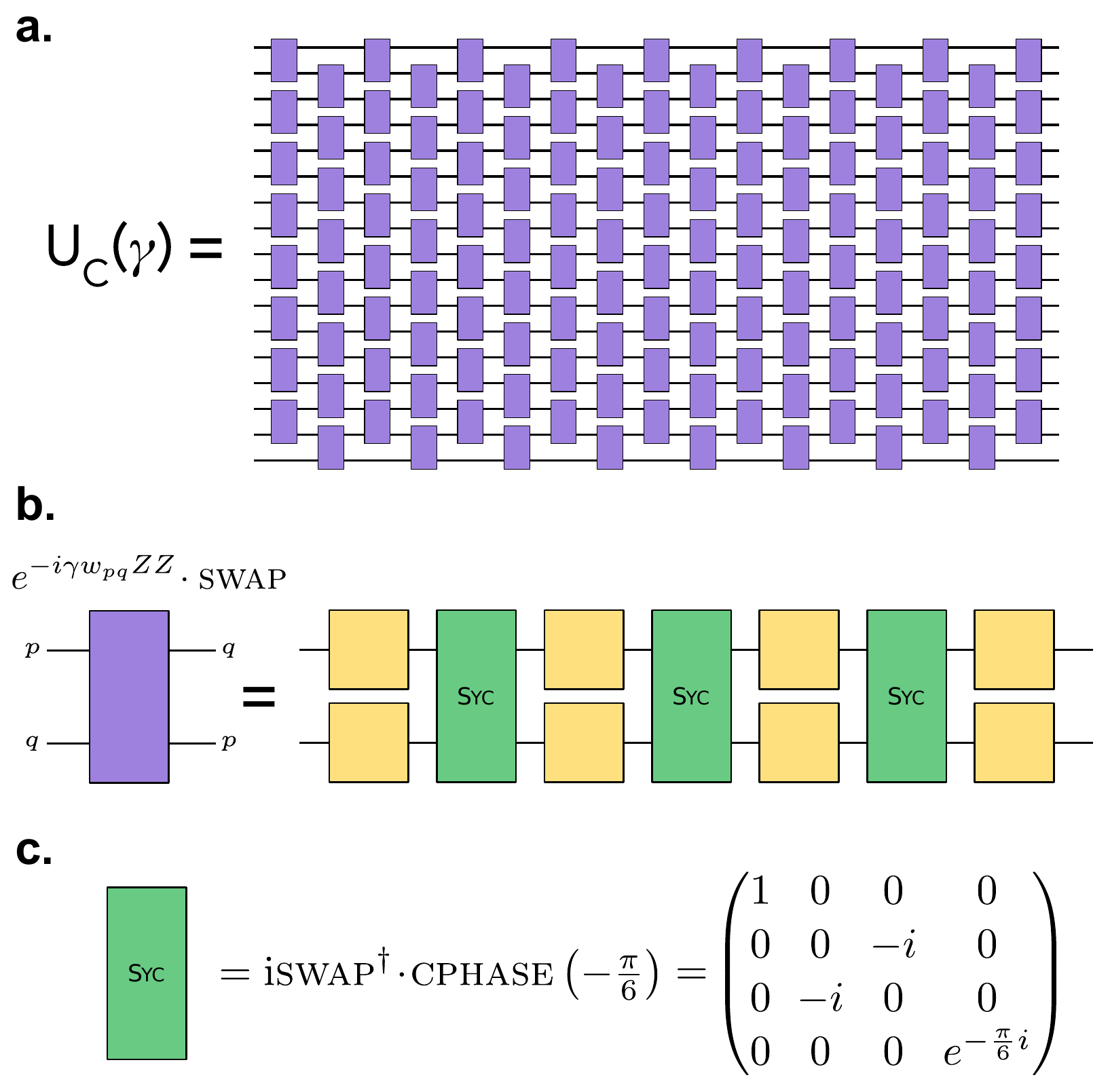}
    \caption{
    \textbf{a.} The linear swap network can route a 17-qubit SK model problem unitary to $n$ layers of nearest-neighbor two qubit interactions. \textbf{b.} The $\zzswap$ interaction is a composite phasing and $\swap$ operation which can be synthesized from three applications of our hardware native entangling $\syc$ and $\gamma w$-dependent single-qubit rotations (yellow boxes). \textbf{c.} The definition of the $\syc$ gate.
    }
    \label{fig:circuit}
\end{figure}

We approach compilation as two distinct steps: routing and gate synthesis. The need for routing arises when simulating $U_C$ for a cost function $C$ defined on a graph that is not a subgraph of our planar hardware connectivity. To simulate such $U_C$ we perform layers of swap gates (forming a swap network) which permute qubits such that all edges in the problem graph correspond to an edge in the hardware graph at least once, at which point the corresponding cost function terms can be implemented. An example of such a swap network is depicted in \figa{circuit}{a}.

The final compilation step, gate synthesis, involves decomposing arbitrary 1- and 2-qubit interactions into physical gates supported by the device (see, e.g.~\figa{circuit}{b}). The physical gates used in this experiment are arbitrary single-qubit rotations and a two-qubit entangling gate native to the Sycamore hardware which we refer to as the $\syc$ gate and define in \figa{circuit}{c}. Through multiple applications of this gate and single-qubit rotations, we are able to realize arbitrary entangling gates. Compilation details can be found in \app{hardware-and-compilation}. The average two-qubit gate fidelities on this device were 99.4\% as measured by cross entropy benchmarking \cite{google_supremacy_2019} and average readout fidelity was 95.9\% per qubit. We now discuss compilation for the three families of optimization problems studied in this work.

\textbf{Hardware Grid Problems.} Swap networks are not required when the problem graph matches the connectivity of our hardware; this is the main reason for studying such problems despite results showing that problems on such graphs are efficient to solve on average \cite{barahona_ising_1982,ronnow_defining_2014}. We generated random instances of hardware grid problems by sampling $w_{ij}$ to be $\pm 1$ for edges in the device topology (and zero otherwise). Gates are scheduled so that the degree-four interaction graph can be implemented in four rounds of two-qubit gates by cycling through the interactions to the left, right, top and bottom of each interior qubit. Each two-qubit $ZZ$ interaction can be synthesized with two layers of hardware-native $\syc$ gates interleaved with $\gamma$-dependent single-qubit rotations. In total, each application of the problem unitary is effected with eight total layers of $\syc$ gates.

\textbf{Sherrington-Kirkpatrick (SK) Model.}
A canonical example of a frustrated spin glass is the Sherrington-Kirkpatrick model \cite{sherrington_kirkpatrick_model_1975}. It is defined on the complete graph with $w_{ij}$ randomly chosen to be $\pm 1$. For large $n$, optimal parameters are independent of the instance \cite{farhi-sk-qaoa}. The SK model is the most challenging model to implement owing to its fully-connected interaction graph. Optimal routing can be performed using the linear swap networks discussed in Ref.~\cite{hirata2011efficient} and depicted in \fig{circuit}. This requires $n$ layers of the composite $\zzswap$ interaction, each of which can be synthesized from three $\syc$ gates with interleaved $\gamma$-dependent single-qubit rotations. Thus, one application of the problem unitary can be effected in $3n$ layers of $\syc$ gates.

\textbf{MaxCut on 3-Regular Graphs.}
MaxCut is a widely studied problem,
and there is a polynomial-time algorithm due to \citet{gw-maxcut} which guarantees a certain approximation ratio for all graphs, and it is an open question whether QAOA can efficiently achieve this or beat it \cite{2019-bravyi-qaoa}. Unlike the previous two problem families, all edge weights are set to $1$, and we sample random 3-regular graphs to generate various instances. The connectivity of the problem Hamiltonian's graph differs for each instance. While one could use the fully-connected swap network to route these circuits, this is wasteful.
Instead, we used the routing functionality from the $\mathrm{t|ket\rangle}$ compiler to heuristically insert $\swap$ operations which move logical assignments to be adjacent \cite{tket-qubit-routing}. These compiled circuits are of roughly equal depth to those from a fully-connected swap network, but the number of two qubit operations is roughly quadratically reduced.

\section{Energy Landscapes and Optimization}\label{sec:optimization}

\begin{figure}[htbp]
    \centering
    \includegraphics[width=0.5\textwidth]{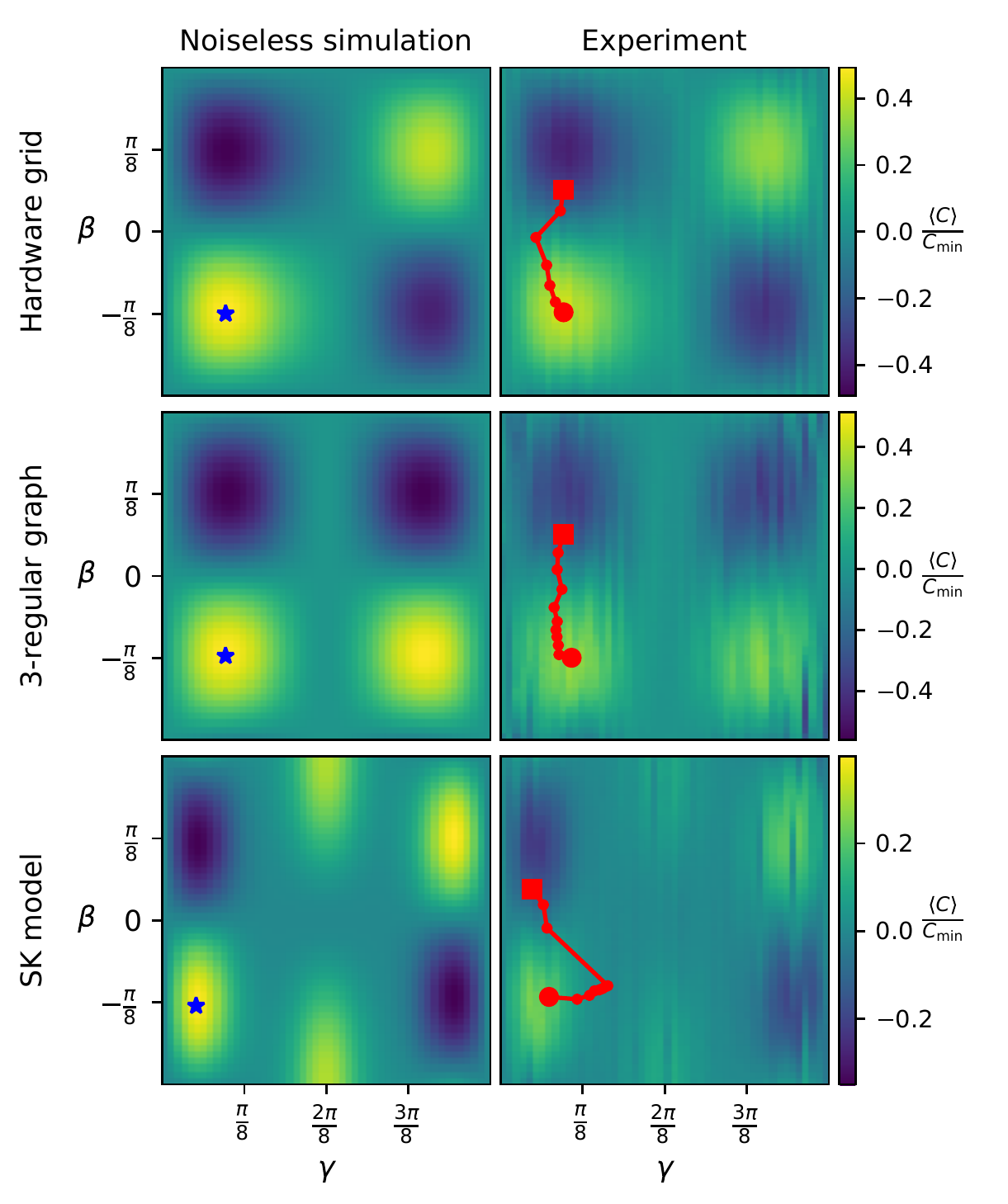}
    \caption{
    Comparison of simulated (left) and experimental (right) $p=1$ landscapes, with a clear correspondence of landscape features. An overlaid optimization trace (red, initialized from square marker) demonstrates the ability of a classical optimizer to find optimal parameters. The blue star in each noiseless plot indicates the theoretical local optimum. Problem sizes are $n = 23$, $n = 14$, and $n = 11$ for Hardware Grid, 3-regular MaxCut, and SK model, respectively.
}
    \label{fig:landscapes}
\end{figure}

QAOA is a variational quantum algorithm where circuit parameters $(\gammavector, \betavector)$ are optimized using a classical optimizer, but function evaluations are executed on a quantum processor \cite{wecker_training_2016, pontryagin-angle-optimization, barren-plateaus}.
First, one repeatedly constructs the state $\ket{\gammavector, \betavector}$ with fixed parameters and samples bitstrings to estimate $\C \equiv \langle \gammavector, \betavector | C | \gammavector, \betavector \rangle$. On our superconducting qubit platform we can sample roughly five thousand bitstrings per second. A classical ``outer-loop" optimizer can then suggest new parameters to decrease the observed expectation value. Note that we normalize by the cost function's true minimum, so we are in fact maximizing $\covercmin$ ($C_{\rm min}$ is negative and hence, minimizing $\C$ corresponds to maximizing $\covercmin$).

For $p=1$, we can visualize the cost function landscape as a function of the parameters $( \gammavector, \betavector) = (\gamma_1, \beta_1)$ in a three-dimensional plot (where we drop the subscripts and label the axes $\gamma$ and $\beta$). The presence of features like hills and valleys in the landscape gives confidence that a classical optimization can be effective. Comparison of simulated and empirical $p=1$ landscapes is a common qualitative diagnostic for the performance of experiments \cite{2018-bristol-photonics, 2019-monroe-qaoa, 2019-julich-qaoa-on-ibm, rigetti-xy-gate, chalmers-2q-qaoa}. For classical optimization to be successful the quantum computer must provide accurate estimates of $\C$. Otherwise, noise can overwhelm any signal making it difficult for a classical optimizer to improve the parameter estimates. Issues such as decoherence, crosstalk, and systematic errors manifest as differences (e.g., damping or warping) from the ideal landscape.

\fig{landscapes} contains simulated theoretical and experimental landscapes for selected instances of the three problem families evaluated on a grid of $\beta \in [-\pi / 4, \pi/4]$ and $\gamma \in [0, \pi/2]$ parameters with a resolution of 50 points along each linear axis. Each expectation value was estimated using 50,000 circuit repetitions with efficient post-processing to compensate for readout bias (see \app{readout}). The hardware grid problem shows clear features at the maximum size of our study, $n=23$. For the other two problems performance degrades with increasing $n$ and so we show data at $n=14$ for the 3-regular graph problem and $n=11$ for the SK model. We highlight the correspondence between experimental and theoretical landscapes for problems of large size and complexity. Prior experimental demonstrations have presented landscapes for a maximum of $n=20$ on a hardware-native interaction graph \cite{2019-monroe-qaoa} and a maximum of $n=4$ for fully-connected problems like the SK model \cite{rigetti-xy-gate}.

In \fig{landscapes}, we also overlay a trace of the classical optimizer's path through parameter space as a red line. We used a classical optimizer called Model Gradient Descent (MGD) which has been shown numerically to perform
well with a small number of function evaluations by using a quadratic surrogate model of the objective function to estimate the gradient \cite{sung-optimizers}. Details are given in \app{optimizer}. In this example, we initialized the parameter optimization from an intentionally bad parameter setting and observed that MGD was able to enter the vicinity of the optimum in 10 iterations or fewer, with each iteration consisting of six energy evaluations of 25,000 shots each.

\section{Hardware Performance of QAOA}\label{sec:performance}

As the name implies, noisy intermediate-scale quantum (NISQ) processors are noisy devices with high error rates and a variety of error channels. Thus, NISQ circuits are expected to degrade in performance as the number of gates is increased. Here, we study the performance of QAOA as implemented on our quantum processor at different $n$ and $p$ using an application-specific metric: the normalized observed cost function $\covercmin$. A value of $1$ is perfect and $0$ corresponds to the performance we would expect from random guessing. In order to distinguish the effects of noise from the robustness provided by using a classical outer-loop optimizer, here we report results obtained from running circuits at the theoretically optimal $(\betavector, \gammavector)$ values.

\begin{figure}[htbp]
    \centering
    \includegraphics[width=0.485\textwidth]{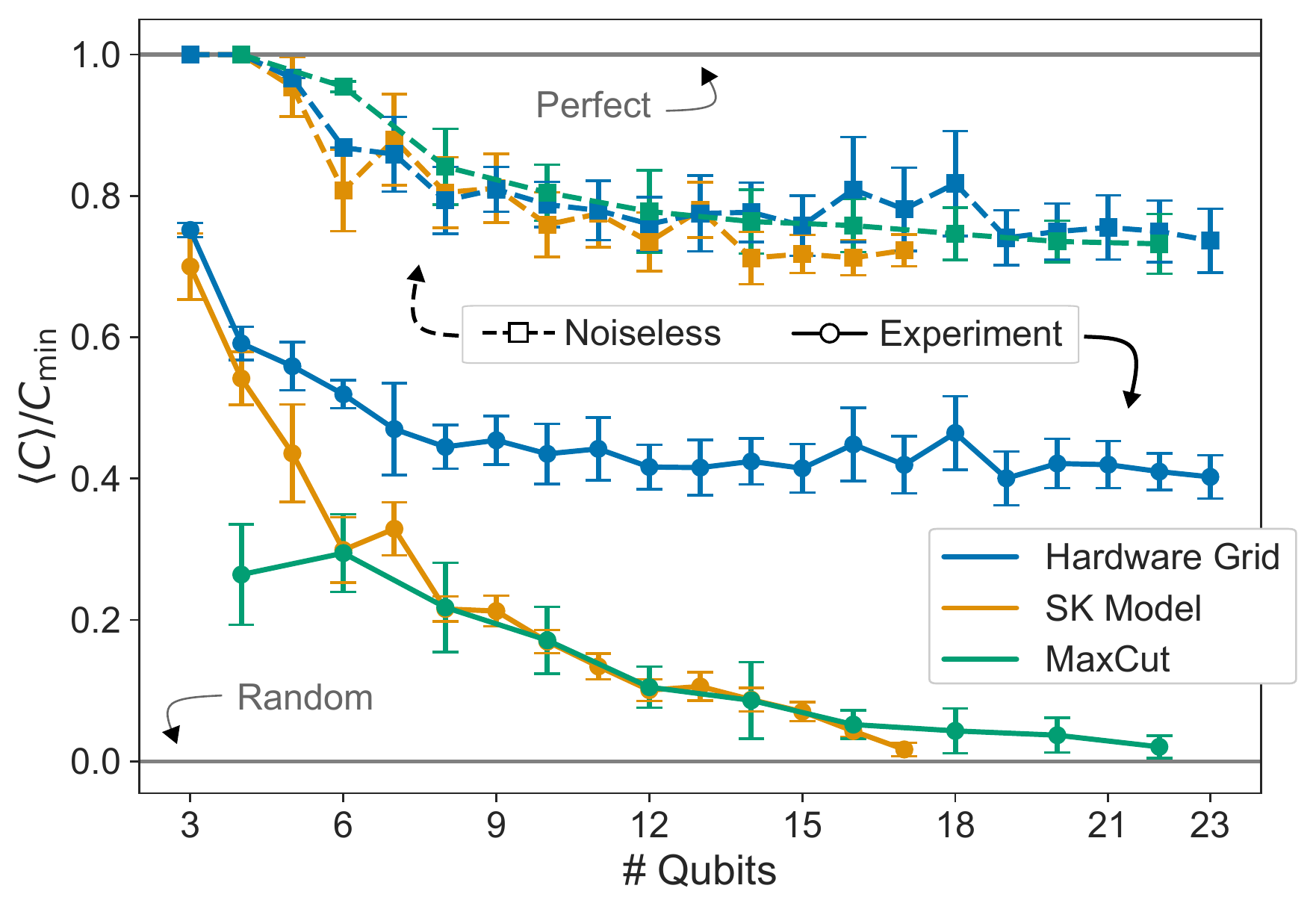}
    \caption{
    QAOA performance as a function of problem size, $n$. Each size is the average over ten random instances (std.~deviation given by error bars). While Hardware Grid problems show $n$-independent noise, we observe that experimental SK model and MaxCut solutions approach those found by random guessing as $n$ is increased.
    }
    \label{fig:expect-val-hardware}
\end{figure}

\begin{figure}[htbp]
    \centering
    \includegraphics[width=0.45\textwidth]{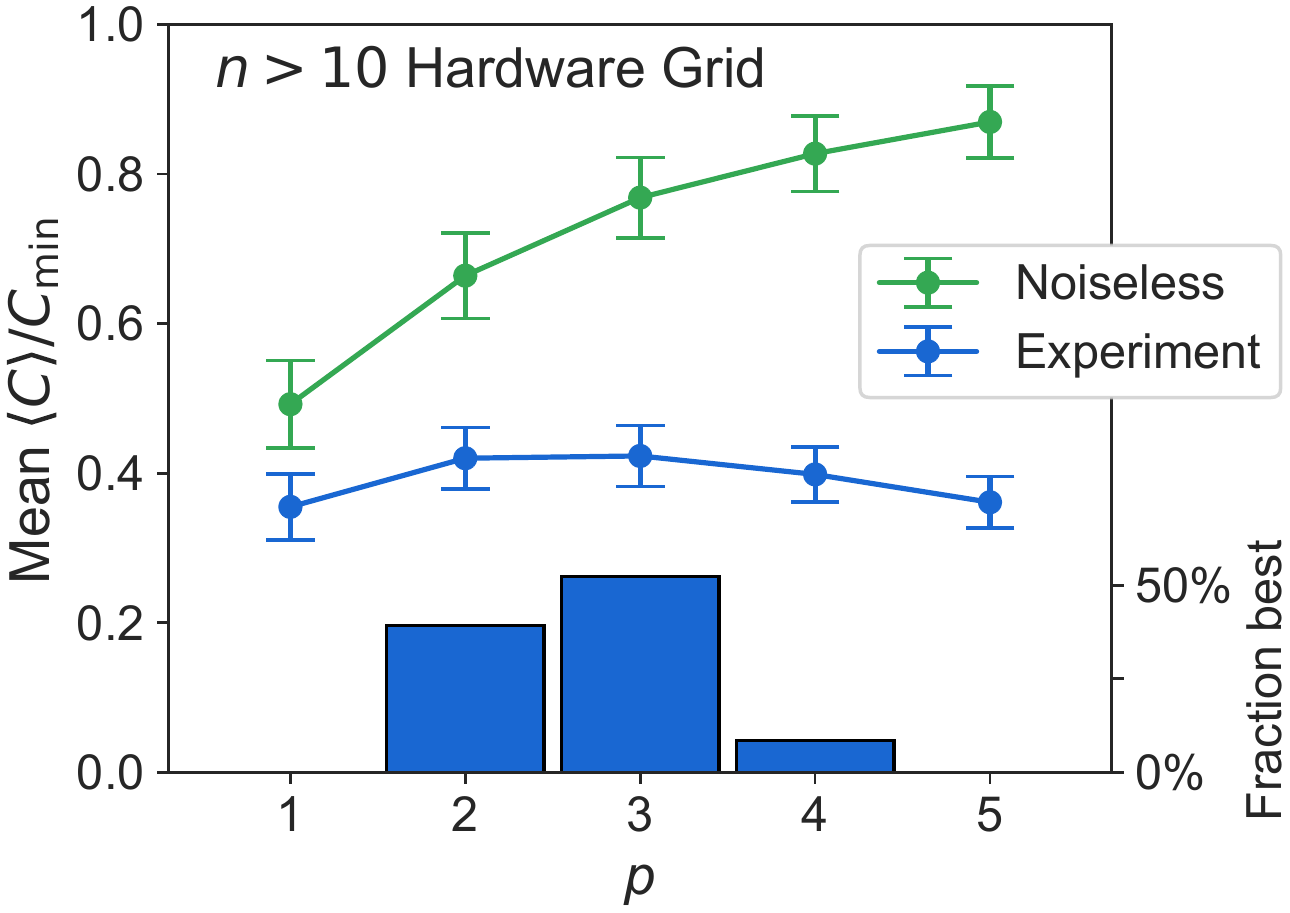}
    \caption{
    QAOA performance as a function of depth, $p$. In ideal simulation, increasing $p$ increases the quality of solutions. For experimental Hardware Grid results, we observe increased performance for $p > 1$ both as measured by the mean over all instances (lines) and statistics of which $p$ maximizes performance on a per-instance basis (histogram). At larger $p$, errors overwhelm the theoretical performance increase.
    }
    \label{fig:hardware-grid-p}
\end{figure}

In \fig{expect-val-hardware}, we observe that $\covercmin$ achieved for the hardware graph seems to saturate to a value that is independent of of $n$. This occurs despite the fact that circuit fidelity is decreasing with increasing $n$. In fact, this is theoretically anticipated behavior that can be understood by moving to the Heisenberg operator formalism and considering an observable $Z_i Z_j$. The expectation value for this operator is conjugated by the circuit unitary involving $p$ applications of the instance graph. This gives an expression for the expectation value of $Z_i Z_j$ which only involves qubits that are at most $p$ edges away from $i$ and $j$. Thus for fixed $p$, the error for a given term is asymptotically unaffected as we grow $n$. Recall that $C$ is a sum of these terms, so the total error scales linearly with $n$; but $\cmin \propto n$ so $\covercmin$ is constant with respect to $n$. Note that non-local error channels or crosstalk could potentially remove this property.

Compiled problems---namely SK model and 3-regular MaxCut problems---result in deeper circuits extensive in the number of qubits. As the depth grows, there is a higher chance of an error occurring. The high degree of the SK model graph and the high effective degree of the MaxCut circuits after compilation means that these errors quickly propagate among all qubits and the quality of solutions can be approximately modeled as the result of a depolarizing channel, with further analysis in \app{all-optimal-angle}.
Even on these challenging problems, we observe performance exceeding random guessing for problem sizes up to 17 bits, even with circuits of depth $p=3$. Note finally that despite circuits with significantly fewer gates (although similar depth), performance on the MaxCut instances tracks performance on the SK model instances rather closely, further substantiating the circuit depth as useful proxy for the performance of QAOA.

In a noiseless case, the quality of a QAOA solution can be improved by increasing the depth parameter $p$. However, the additional depth also increases the probability of error. We study this interplay between noise and algorithmic power in \fig{hardware-grid-p}. Previously, improved performance with $p>1$ had only been experimentally demonstrated for an $n=2$ problem \cite{chalmers-2q-qaoa}. For larger problems ($n=20$), performance for $p=2$ was shown to be within error bars of the $p=1$ performance \cite{2019-monroe-qaoa}. \fig{hardware-grid-p} shows the $p$-dependence averaged across all 130 instances where $n>10$. The mean finds its maximum at $p=3$, although there are variations among the instances comparable in scale to the experimental $p$-dependence.
The relatively flat dependence of performance on depth suggests that the experimental noise seems to nearly balance the increase in theoretical performance for this problem family.
For a more meaningful aggregation of the many random instances across problem sizes,
we consider each instance individually and identify which value of the hyperparameter $p$ maximizes performance for that particular instance.
A histogram of these per-instance maximal values is inset in \fig{hardware-grid-p}, showing that
performance is maximized at $p=3$ for over half of instances larger than ten qubits.
Note finally that our full dataset (see \app{all-optimal-angle}) includes per-instance data at all settings of $p$.

\section{Conclusion}

Discrete optimization is an enticing application for near-term devices owing to both the potential value of solutions as well as the viability of heuristic low-depth algorithms such as the QAOA. While no existing quantum processors can outperform classical optimization heuristics, the application of popular methods such as the QAOA to prototypical problems can be used as a benchmark for comparing various hardware platforms.

Previous demonstrations of the QAOA have primarily optimized problems tailored to the hardware architecture at minimal depth. Using the Google ``Sycamore" platform, we explored these types of problems, which we termed Hardware Grid problems, and demonstrated robust performance at large numbers of qubits. We showed that the locations of maxima and minima in the $p=1$ diagnostic landscape match those from the theoretically computed surface, and that variational optimization can still find the optimum with noisy quantum objective function evaluation. We also applied the QAOA to various problem sizes using pre-computed parameters from noiseless simulation, and observed an $n$-independent noise effect on the approximation ratios for Hardware Grid problems. This is consistent with our theoretical understanding that the noise-induced degradation of each term in the objective function remains constant in the shallow-depth regime where correlations remain local. Furthermore, we report the first clear cases of performance maximization at $p=3$ for the QAOA owing to the low error rate of our hardware.

Most real world instances of combinatorial optimization problems cannot be mapped to hardware-native topologies without significant additional resources. Instead, problems must be compiled by routing qubits with swap networks. This additional overhead can have a significant impact on the algorithm's performance. We studied random instances of the fully-connected SK model. Although we report non-negligible performance for large ($n=17$), deep ($p=3$), and complex (fully-connected) problems, we see that performance degrades with problem size for such instances.

The promise of quantum enhanced optimization will continue to motivate the development of new quantum technology and algorithms. Nevertheless, for quantum optimization to compete with classical methods for real-world problems, it is necessary to push beyond contrived problems at low circuit depth. Our work demonstrates important progress in the implementation and performance of quantum optimization algorithms on a real device, and underscores the challenges in applying these algorithms beyond those natively realized by hardware interaction graphs.\\

\subsection*{Acknowledgements}

The authors thank the Cambridge Quantum Computing team for helpful correspondence about their $\rm{t}\ket{\rm{ket}}$ compiler, which we used for routing of MaxCut problems. The VW team acknowledges support from the European Union's Horizon 2020 research and innovation program under grant agreement No.~828826 ``Quromorphic". We thank all other members of the Google Quantum team, as well as our executive sponsors. Dave Bacon is a CIFAR Associate Fellow in the Quantum Information Science Program.

\subsection*{Author Contributions}

R.~Babbush and E.~Farhi designed the experiment. M.~Harrigan and K.~Sung led code development and data collection with assistance from non-Google collaborators. Z.~Jiang and N.~Rubin derived the gate synthesis used in compilation. The manuscript was written by M.~Harrigan, R.~Babbush, E.~Farhi, and K.~Sung. Experiments were performed using cloud access to a quantum processor that was recently developed and fabricated by a large effort involving the entire Google Quantum team.

\subsection*{Code and Data Availability}

The code used in this experiment is available at \url{https://github.com/quantumlib/ReCirq}.
The experimental data for this experiment is available on Figshare, see Ref.~\citenum{qaoadata}.

\bibliography{qaoa}

\begin{thebibliography}{39}%
\makeatletter
\providecommand \@ifxundefined [1]{%
 \@ifx{#1\undefined}
}%
\providecommand \@ifnum [1]{%
 \ifnum #1\expandafter \@firstoftwo
 \else \expandafter \@secondoftwo
 \fi
}%
\providecommand \@ifx [1]{%
 \ifx #1\expandafter \@firstoftwo
 \else \expandafter \@secondoftwo
 \fi
}%
\providecommand \natexlab [1]{#1}%
\providecommand \enquote  [1]{``#1''}%
\providecommand \bibnamefont  [1]{#1}%
\providecommand \bibfnamefont [1]{#1}%
\providecommand \citenamefont [1]{#1}%
\providecommand \href@noop [0]{\@secondoftwo}%
\providecommand \href [0]{\begingroup \@sanitize@url \@href}%
\providecommand \@href[1]{\@@startlink{#1}\@@href}%
\providecommand \@@href[1]{\endgroup#1\@@endlink}%
\providecommand \@sanitize@url [0]{\catcode `\\12\catcode `\$12\catcode
  `\&12\catcode `\#12\catcode `\^12\catcode `\_12\catcode `\%12\relax}%
\providecommand \@@startlink[1]{}%
\providecommand \@@endlink[0]{}%
\providecommand \url  [0]{\begingroup\@sanitize@url \@url }%
\providecommand \@url [1]{\endgroup\@href {#1}{\urlprefix }}%
\providecommand \urlprefix  [0]{URL }%
\providecommand \Eprint [0]{\href }%
\providecommand \doibase [0]{http://dx.doi.org/}%
\providecommand \selectlanguage [0]{\@gobble}%
\providecommand \bibinfo  [0]{\@secondoftwo}%
\providecommand \bibfield  [0]{\@secondoftwo}%
\providecommand \translation [1]{[#1]}%
\providecommand \BibitemOpen [0]{}%
\providecommand \bibitemStop [0]{}%
\providecommand \bibitemNoStop [0]{.\EOS\space}%
\providecommand \EOS [0]{\spacefactor3000\relax}%
\providecommand \BibitemShut  [1]{\csname bibitem#1\endcsname}%
\let\auto@bib@innerbib\@empty
\bibitem [{\citenamefont {Arute}\ \emph {et~al.}(2019)\citenamefont {Arute},
  \citenamefont {Arya}, \citenamefont {Babbush}, \citenamefont {Bacon},
  \citenamefont {Bardin}, \citenamefont {Barends}, \citenamefont {Biswas},
  \citenamefont {Boixo}, \citenamefont {Brandao}, \citenamefont {Buell},
  \citenamefont {Burkett}, \citenamefont {Chen}, \citenamefont {Chen},
  \citenamefont {Chiaro}, \citenamefont {Collins}, \citenamefont {Courtney},
  \citenamefont {Dunsworth}, \citenamefont {Farhi}, \citenamefont {Foxen},
  \citenamefont {Fowler}, \citenamefont {Gidney}, \citenamefont {Giustina},
  \citenamefont {Graff}, \citenamefont {Guerin}, \citenamefont {Habegger},
  \citenamefont {Harrigan}, \citenamefont {Hartmann}, \citenamefont {Ho},
  \citenamefont {Hoffmann}, \citenamefont {Huang}, \citenamefont {Humble},
  \citenamefont {Isakov}, \citenamefont {Jeffrey}, \citenamefont {Jiang},
  \citenamefont {Kafri}, \citenamefont {Kechedzhi}, \citenamefont {Kelly},
  \citenamefont {Klimov}, \citenamefont {Knysh}, \citenamefont {Korotkov},
  \citenamefont {Kostritsa}, \citenamefont {Landhuis}, \citenamefont
  {Lindmark}, \citenamefont {Lucero}, \citenamefont {Lyakh}, \citenamefont
  {Mandr{\`a}}, \citenamefont {McClean}, \citenamefont {McEwen}, \citenamefont
  {Megrant}, \citenamefont {Mi}, \citenamefont {Michielsen}, \citenamefont
  {Mohseni}, \citenamefont {Mutus}, \citenamefont {Naaman}, \citenamefont
  {Neeley}, \citenamefont {Neill}, \citenamefont {Niu}, \citenamefont {Ostby},
  \citenamefont {Petukhov}, \citenamefont {Platt}, \citenamefont {Quintana},
  \citenamefont {Rieffel}, \citenamefont {Roushan}, \citenamefont {Rubin},
  \citenamefont {Sank}, \citenamefont {Satzinger}, \citenamefont {Smelyanskiy},
  \citenamefont {Sung}, \citenamefont {Trevithick}, \citenamefont
  {Vainsencher}, \citenamefont {Villalonga}, \citenamefont {White},
  \citenamefont {Yao}, \citenamefont {Yeh}, \citenamefont {Zalcman},
  \citenamefont {Neven},\ and\ \citenamefont
  {Martinis}}]{google_supremacy_2019}%
  \BibitemOpen
  \bibfield  {author} {\bibinfo {author} {\bibfnamefont {Frank}\ \bibnamefont
  {Arute}}, \bibinfo {author} {\bibfnamefont {Kunal}\ \bibnamefont {Arya}},
  \bibinfo {author} {\bibfnamefont {Ryan}\ \bibnamefont {Babbush}}, \bibinfo
  {author} {\bibfnamefont {Dave}\ \bibnamefont {Bacon}}, \bibinfo {author}
  {\bibfnamefont {Joseph~C.}\ \bibnamefont {Bardin}}, \bibinfo {author}
  {\bibfnamefont {Rami}\ \bibnamefont {Barends}}, \bibinfo {author}
  {\bibfnamefont {Rupak}\ \bibnamefont {Biswas}}, \bibinfo {author}
  {\bibfnamefont {Sergio}\ \bibnamefont {Boixo}}, \bibinfo {author}
  {\bibfnamefont {Fernando G. S.~L.}\ \bibnamefont {Brandao}}, \bibinfo
  {author} {\bibfnamefont {David~A.}\ \bibnamefont {Buell}}, \bibinfo {author}
  {\bibfnamefont {Brian}\ \bibnamefont {Burkett}}, \bibinfo {author}
  {\bibfnamefont {Yu}~\bibnamefont {Chen}}, \bibinfo {author} {\bibfnamefont
  {Zijun}\ \bibnamefont {Chen}}, \bibinfo {author} {\bibfnamefont {Ben}\
  \bibnamefont {Chiaro}}, \bibinfo {author} {\bibfnamefont {Roberto}\
  \bibnamefont {Collins}}, \bibinfo {author} {\bibfnamefont {William}\
  \bibnamefont {Courtney}}, \bibinfo {author} {\bibfnamefont {Andrew}\
  \bibnamefont {Dunsworth}}, \bibinfo {author} {\bibfnamefont {Edward}\
  \bibnamefont {Farhi}}, \bibinfo {author} {\bibfnamefont {Brooks}\
  \bibnamefont {Foxen}}, \bibinfo {author} {\bibfnamefont {Austin}\
  \bibnamefont {Fowler}}, \bibinfo {author} {\bibfnamefont {Craig}\
  \bibnamefont {Gidney}}, \bibinfo {author} {\bibfnamefont {Marissa}\
  \bibnamefont {Giustina}}, \bibinfo {author} {\bibfnamefont {Rob}\
  \bibnamefont {Graff}}, \bibinfo {author} {\bibfnamefont {Keith}\ \bibnamefont
  {Guerin}}, \bibinfo {author} {\bibfnamefont {Steve}\ \bibnamefont
  {Habegger}}, \bibinfo {author} {\bibfnamefont {Matthew~P.}\ \bibnamefont
  {Harrigan}}, \bibinfo {author} {\bibfnamefont {Michael~J.}\ \bibnamefont
  {Hartmann}}, \bibinfo {author} {\bibfnamefont {Alan}\ \bibnamefont {Ho}},
  \bibinfo {author} {\bibfnamefont {Markus}\ \bibnamefont {Hoffmann}}, \bibinfo
  {author} {\bibfnamefont {Trent}\ \bibnamefont {Huang}}, \bibinfo {author}
  {\bibfnamefont {Travis~S.}\ \bibnamefont {Humble}}, \bibinfo {author}
  {\bibfnamefont {Sergei~V.}\ \bibnamefont {Isakov}}, \bibinfo {author}
  {\bibfnamefont {Evan}\ \bibnamefont {Jeffrey}}, \bibinfo {author}
  {\bibfnamefont {Zhang}\ \bibnamefont {Jiang}}, \bibinfo {author}
  {\bibfnamefont {Dvir}\ \bibnamefont {Kafri}}, \bibinfo {author}
  {\bibfnamefont {Kostyantyn}\ \bibnamefont {Kechedzhi}}, \bibinfo {author}
  {\bibfnamefont {Julian}\ \bibnamefont {Kelly}}, \bibinfo {author}
  {\bibfnamefont {Paul~V.}\ \bibnamefont {Klimov}}, \bibinfo {author}
  {\bibfnamefont {Sergey}\ \bibnamefont {Knysh}}, \bibinfo {author}
  {\bibfnamefont {Alexander}\ \bibnamefont {Korotkov}}, \bibinfo {author}
  {\bibfnamefont {Fedor}\ \bibnamefont {Kostritsa}}, \bibinfo {author}
  {\bibfnamefont {David}\ \bibnamefont {Landhuis}}, \bibinfo {author}
  {\bibfnamefont {Mike}\ \bibnamefont {Lindmark}}, \bibinfo {author}
  {\bibfnamefont {Erik}\ \bibnamefont {Lucero}}, \bibinfo {author}
  {\bibfnamefont {Dmitry}\ \bibnamefont {Lyakh}}, \bibinfo {author}
  {\bibfnamefont {Salvatore}\ \bibnamefont {Mandr{\`a}}}, \bibinfo {author}
  {\bibfnamefont {Jarrod~R.}\ \bibnamefont {McClean}}, \bibinfo {author}
  {\bibfnamefont {Matthew}\ \bibnamefont {McEwen}}, \bibinfo {author}
  {\bibfnamefont {Anthony}\ \bibnamefont {Megrant}}, \bibinfo {author}
  {\bibfnamefont {Xiao}\ \bibnamefont {Mi}}, \bibinfo {author} {\bibfnamefont
  {Kristel}\ \bibnamefont {Michielsen}}, \bibinfo {author} {\bibfnamefont
  {Masoud}\ \bibnamefont {Mohseni}}, \bibinfo {author} {\bibfnamefont {Josh}\
  \bibnamefont {Mutus}}, \bibinfo {author} {\bibfnamefont {Ofer}\ \bibnamefont
  {Naaman}}, \bibinfo {author} {\bibfnamefont {Matthew}\ \bibnamefont
  {Neeley}}, \bibinfo {author} {\bibfnamefont {Charles}\ \bibnamefont {Neill}},
  \bibinfo {author} {\bibfnamefont {Murphy~Yuezhen}\ \bibnamefont {Niu}},
  \bibinfo {author} {\bibfnamefont {Eric}\ \bibnamefont {Ostby}}, \bibinfo
  {author} {\bibfnamefont {Andre}\ \bibnamefont {Petukhov}}, \bibinfo {author}
  {\bibfnamefont {John~C.}\ \bibnamefont {Platt}}, \bibinfo {author}
  {\bibfnamefont {Chris}\ \bibnamefont {Quintana}}, \bibinfo {author}
  {\bibfnamefont {Eleanor~G.}\ \bibnamefont {Rieffel}}, \bibinfo {author}
  {\bibfnamefont {Pedram}\ \bibnamefont {Roushan}}, \bibinfo {author}
  {\bibfnamefont {Nicholas~C.}\ \bibnamefont {Rubin}}, \bibinfo {author}
  {\bibfnamefont {Daniel}\ \bibnamefont {Sank}}, \bibinfo {author}
  {\bibfnamefont {Kevin~J.}\ \bibnamefont {Satzinger}}, \bibinfo {author}
  {\bibfnamefont {Vadim}\ \bibnamefont {Smelyanskiy}}, \bibinfo {author}
  {\bibfnamefont {Kevin~J.}\ \bibnamefont {Sung}}, \bibinfo {author}
  {\bibfnamefont {Matthew~D.}\ \bibnamefont {Trevithick}}, \bibinfo {author}
  {\bibfnamefont {Amit}\ \bibnamefont {Vainsencher}}, \bibinfo {author}
  {\bibfnamefont {Benjamin}\ \bibnamefont {Villalonga}}, \bibinfo {author}
  {\bibfnamefont {Theodore}\ \bibnamefont {White}}, \bibinfo {author}
  {\bibfnamefont {Z.~Jamie}\ \bibnamefont {Yao}}, \bibinfo {author}
  {\bibfnamefont {Ping}\ \bibnamefont {Yeh}}, \bibinfo {author} {\bibfnamefont
  {Adam}\ \bibnamefont {Zalcman}}, \bibinfo {author} {\bibfnamefont {Hartmut}\
  \bibnamefont {Neven}}, \ and\ \bibinfo {author} {\bibfnamefont {John~M.}\
  \bibnamefont {Martinis}},\ }\bibfield  {title} {\enquote {\bibinfo {title}
  {Quantum supremacy using a programmable superconducting processor},}\ }\href
  {\doibase 10.1038/s41586-019-1666-5} {\bibfield  {journal} {\bibinfo
  {journal} {Nature}\ }\textbf {\bibinfo {volume} {574}},\ \bibinfo {pages}
  {505--510} (\bibinfo {year} {2019})}\BibitemShut {NoStop}%
\bibitem [{\citenamefont {Aspuru-Guzik}\ \emph {et~al.}(2005)\citenamefont
  {Aspuru-Guzik}, \citenamefont {Dutoi}, \citenamefont {Love},\ and\
  \citenamefont {Head-Gordon}}]{aspuru_guzik_simulated_2005}%
  \BibitemOpen
  \bibfield  {author} {\bibinfo {author} {\bibfnamefont {Al{\'a}n}\
  \bibnamefont {Aspuru-Guzik}}, \bibinfo {author} {\bibfnamefont {Anthony~D.}\
  \bibnamefont {Dutoi}}, \bibinfo {author} {\bibfnamefont {Peter~J.}\
  \bibnamefont {Love}}, \ and\ \bibinfo {author} {\bibfnamefont {Martin}\
  \bibnamefont {Head-Gordon}},\ }\bibfield  {title} {\enquote {\bibinfo {title}
  {Simulated quantum computation of molecular energies},}\ }\href {\doibase
  10.1126/science.1113479} {\bibfield  {journal} {\bibinfo  {journal}
  {Science}\ }\textbf {\bibinfo {volume} {309}},\ \bibinfo {pages} {1704--1707}
  (\bibinfo {year} {2005})}\BibitemShut {NoStop}%
\bibitem [{\citenamefont {O'Malley}\ \emph {et~al.}(2016)\citenamefont
  {O'Malley}, \citenamefont {Babbush}, \citenamefont {Kivlichan}, \citenamefont
  {Romero}, \citenamefont {McClean}, \citenamefont {Barends}, \citenamefont
  {Kelly}, \citenamefont {Roushan}, \citenamefont {Tranter}, \citenamefont
  {Ding}, \citenamefont {Campbell}, \citenamefont {Chen}, \citenamefont {Chen},
  \citenamefont {Chiaro}, \citenamefont {Dunsworth}, \citenamefont {Fowler},
  \citenamefont {Jeffrey}, \citenamefont {Lucero}, \citenamefont {Megrant},
  \citenamefont {Mutus}, \citenamefont {Neeley}, \citenamefont {Neill},
  \citenamefont {Quintana}, \citenamefont {Sank}, \citenamefont {Vainsencher},
  \citenamefont {Wenner}, \citenamefont {White}, \citenamefont {Coveney},
  \citenamefont {Love}, \citenamefont {Neven}, \citenamefont {Aspuru-Guzik},\
  and\ \citenamefont {Martinis}}]{omalley_scalable_2016}%
  \BibitemOpen
  \bibfield  {author} {\bibinfo {author} {\bibfnamefont {P.~J.~J.}\
  \bibnamefont {O'Malley}}, \bibinfo {author} {\bibfnamefont {R.}~\bibnamefont
  {Babbush}}, \bibinfo {author} {\bibfnamefont {I.~D.}\ \bibnamefont
  {Kivlichan}}, \bibinfo {author} {\bibfnamefont {J.}~\bibnamefont {Romero}},
  \bibinfo {author} {\bibfnamefont {J.~R.}\ \bibnamefont {McClean}}, \bibinfo
  {author} {\bibfnamefont {R.}~\bibnamefont {Barends}}, \bibinfo {author}
  {\bibfnamefont {J.}~\bibnamefont {Kelly}}, \bibinfo {author} {\bibfnamefont
  {P.}~\bibnamefont {Roushan}}, \bibinfo {author} {\bibfnamefont
  {A.}~\bibnamefont {Tranter}}, \bibinfo {author} {\bibfnamefont
  {N.}~\bibnamefont {Ding}}, \bibinfo {author} {\bibfnamefont {B.}~\bibnamefont
  {Campbell}}, \bibinfo {author} {\bibfnamefont {Y.}~\bibnamefont {Chen}},
  \bibinfo {author} {\bibfnamefont {Z.}~\bibnamefont {Chen}}, \bibinfo {author}
  {\bibfnamefont {B.}~\bibnamefont {Chiaro}}, \bibinfo {author} {\bibfnamefont
  {A.}~\bibnamefont {Dunsworth}}, \bibinfo {author} {\bibfnamefont {A.~G.}\
  \bibnamefont {Fowler}}, \bibinfo {author} {\bibfnamefont {E.}~\bibnamefont
  {Jeffrey}}, \bibinfo {author} {\bibfnamefont {E.}~\bibnamefont {Lucero}},
  \bibinfo {author} {\bibfnamefont {A.}~\bibnamefont {Megrant}}, \bibinfo
  {author} {\bibfnamefont {J.~Y.}\ \bibnamefont {Mutus}}, \bibinfo {author}
  {\bibfnamefont {M.}~\bibnamefont {Neeley}}, \bibinfo {author} {\bibfnamefont
  {C.}~\bibnamefont {Neill}}, \bibinfo {author} {\bibfnamefont
  {C.}~\bibnamefont {Quintana}}, \bibinfo {author} {\bibfnamefont
  {D.}~\bibnamefont {Sank}}, \bibinfo {author} {\bibfnamefont {A.}~\bibnamefont
  {Vainsencher}}, \bibinfo {author} {\bibfnamefont {J.}~\bibnamefont {Wenner}},
  \bibinfo {author} {\bibfnamefont {T.~C.}\ \bibnamefont {White}}, \bibinfo
  {author} {\bibfnamefont {P.~V.}\ \bibnamefont {Coveney}}, \bibinfo {author}
  {\bibfnamefont {P.~J.}\ \bibnamefont {Love}}, \bibinfo {author}
  {\bibfnamefont {H.}~\bibnamefont {Neven}}, \bibinfo {author} {\bibfnamefont
  {A.}~\bibnamefont {Aspuru-Guzik}}, \ and\ \bibinfo {author} {\bibfnamefont
  {J.~M.}\ \bibnamefont {Martinis}},\ }\bibfield  {title} {\enquote {\bibinfo
  {title} {Scalable quantum simulation of molecular energies},}\ }\href
  {\doibase 10.1103/PhysRevX.6.031007} {\bibfield  {journal} {\bibinfo
  {journal} {Phys. Rev. X}\ }\textbf {\bibinfo {volume} {6}},\ \bibinfo {pages}
  {031007} (\bibinfo {year} {2016})}\BibitemShut {NoStop}%
\bibitem [{\citenamefont {Biamonte}\ \emph {et~al.}(2016)\citenamefont
  {Biamonte}, \citenamefont {Wittek}, \citenamefont {Pancotti}, \citenamefont
  {Rebentrost}, \citenamefont {Wiebe},\ and\ \citenamefont
  {Lloyd}}]{2017-biamonte-quantum-machine-learning}%
  \BibitemOpen
  \bibfield  {author} {\bibinfo {author} {\bibfnamefont {Jacob}\ \bibnamefont
  {Biamonte}}, \bibinfo {author} {\bibfnamefont {Peter}\ \bibnamefont
  {Wittek}}, \bibinfo {author} {\bibfnamefont {Nicola}\ \bibnamefont
  {Pancotti}}, \bibinfo {author} {\bibfnamefont {Patrick}\ \bibnamefont
  {Rebentrost}}, \bibinfo {author} {\bibfnamefont {Nathan}\ \bibnamefont
  {Wiebe}}, \ and\ \bibinfo {author} {\bibfnamefont {Seth}\ \bibnamefont
  {Lloyd}},\ }\bibfield  {title} {\enquote {\bibinfo {title} {Quantum machine
  learning},}\ }\href {\doibase 10.1038/nature23474} {\bibfield  {journal}
  {\bibinfo  {journal} {Nature}\ }\textbf {\bibinfo {volume} {549}},\ \bibinfo
  {pages} {195--202} (\bibinfo {year} {2016})}\BibitemShut {NoStop}%
\bibitem [{\citenamefont {Lloyd}(1996)}]{lloyd_universal_1996}%
  \BibitemOpen
  \bibfield  {author} {\bibinfo {author} {\bibfnamefont {Seth}\ \bibnamefont
  {Lloyd}},\ }\bibfield  {title} {\enquote {\bibinfo {title} {Universal quantum
  simulators},}\ }\href {\doibase 10.1126/science.273.5278.1073} {\bibfield
  {journal} {\bibinfo  {journal} {Science}\ }\textbf {\bibinfo {volume}
  {273}},\ \bibinfo {pages} {1073--1078} (\bibinfo {year} {1996})}\BibitemShut
  {NoStop}%
\bibitem [{\citenamefont {Kadowaki}\ and\ \citenamefont
  {Nishimori}(1998)}]{kadowaki_annealing_1998}%
  \BibitemOpen
  \bibfield  {author} {\bibinfo {author} {\bibfnamefont {Tadashi}\ \bibnamefont
  {Kadowaki}}\ and\ \bibinfo {author} {\bibfnamefont {Hidetoshi}\ \bibnamefont
  {Nishimori}},\ }\bibfield  {title} {\enquote {\bibinfo {title} {Quantum
  annealing in the transverse {Ising} model},}\ }\href {\doibase
  10.1103/PhysRevE.58.5355} {\bibfield  {journal} {\bibinfo  {journal} {Phys.
  Rev. E}\ }\textbf {\bibinfo {volume} {58}},\ \bibinfo {pages} {5355--5363}
  (\bibinfo {year} {1998})}\BibitemShut {NoStop}%
\bibitem [{\citenamefont {Farhi}\ \emph {et~al.}(2001)\citenamefont {Farhi},
  \citenamefont {Goldstone}, \citenamefont {Gutmann}, \citenamefont {Lapan},
  \citenamefont {Lundgren},\ and\ \citenamefont {Preda}}]{farhi_applied_2001}%
  \BibitemOpen
  \bibfield  {author} {\bibinfo {author} {\bibfnamefont {Edward}\ \bibnamefont
  {Farhi}}, \bibinfo {author} {\bibfnamefont {Jeffrey}\ \bibnamefont
  {Goldstone}}, \bibinfo {author} {\bibfnamefont {Sam}\ \bibnamefont
  {Gutmann}}, \bibinfo {author} {\bibfnamefont {Joshua}\ \bibnamefont {Lapan}},
  \bibinfo {author} {\bibfnamefont {Andrew}\ \bibnamefont {Lundgren}}, \ and\
  \bibinfo {author} {\bibfnamefont {Daniel}\ \bibnamefont {Preda}},\ }\bibfield
   {title} {\enquote {\bibinfo {title} {A quantum adiabatic evolution algorithm
  applied to random instances of an {NP}-complete problem},}\ }\href {\doibase
  10.1126/science.1057726} {\bibfield  {journal} {\bibinfo  {journal}
  {Science}\ }\textbf {\bibinfo {volume} {292}},\ \bibinfo {pages} {472--475}
  (\bibinfo {year} {2001})}\BibitemShut {NoStop}%
\bibitem [{\citenamefont {Lucas}(2014)}]{lucas-ising}%
  \BibitemOpen
  \bibfield  {author} {\bibinfo {author} {\bibfnamefont {Andrew}\ \bibnamefont
  {Lucas}},\ }\bibfield  {title} {\enquote {\bibinfo {title} {Ising
  formulations of many {NP} problems},}\ }\href {\doibase
  10.3389/fphy.2014.00005} {\bibfield  {journal} {\bibinfo  {journal}
  {Frontiers in Physics}\ }\textbf {\bibinfo {volume} {2}},\ \bibinfo {pages}
  {5} (\bibinfo {year} {2014})}\BibitemShut {NoStop}%
\bibitem [{\citenamefont {Barahona}(1982)}]{barahona_ising_1982}%
  \BibitemOpen
  \bibfield  {author} {\bibinfo {author} {\bibfnamefont {F}~\bibnamefont
  {Barahona}},\ }\bibfield  {title} {\enquote {\bibinfo {title} {On the
  computational complexity of {Ising} spin glass models},}\ }\href {\doibase
  10.1088/0305-4470/15/10/028} {\bibfield  {journal} {\bibinfo  {journal}
  {Journal of Physics A: Mathematical and General}\ }\textbf {\bibinfo {volume}
  {15}},\ \bibinfo {pages} {3241--3253} (\bibinfo {year} {1982})}\BibitemShut
  {NoStop}%
\bibitem [{\citenamefont {Choi}(2008)}]{choi_minor_2008}%
  \BibitemOpen
  \bibfield  {author} {\bibinfo {author} {\bibfnamefont {Vicky}\ \bibnamefont
  {Choi}},\ }\bibfield  {title} {\enquote {\bibinfo {title} {Minor-embedding in
  adiabatic quantum computation: I. the parameter setting problem},}\ }\href
  {\doibase 10.1007/s11128-008-0082-9} {\bibfield  {journal} {\bibinfo
  {journal} {Quantum Information Processing}\ }\textbf {\bibinfo {volume}
  {7}},\ \bibinfo {pages} {193--209} (\bibinfo {year} {2008})}\BibitemShut
  {NoStop}%
\bibitem [{\citenamefont {Denchev}\ \emph {et~al.}(2016)\citenamefont
  {Denchev}, \citenamefont {Boixo}, \citenamefont {Isakov}, \citenamefont
  {Ding}, \citenamefont {Babbush}, \citenamefont {Smelyanskiy}, \citenamefont
  {Martinis},\ and\ \citenamefont {Neven}}]{denchev_tunneling_2016}%
  \BibitemOpen
  \bibfield  {author} {\bibinfo {author} {\bibfnamefont {Vasil~S.}\
  \bibnamefont {Denchev}}, \bibinfo {author} {\bibfnamefont {Sergio}\
  \bibnamefont {Boixo}}, \bibinfo {author} {\bibfnamefont {Sergei~V.}\
  \bibnamefont {Isakov}}, \bibinfo {author} {\bibfnamefont {Nan}\ \bibnamefont
  {Ding}}, \bibinfo {author} {\bibfnamefont {Ryan}\ \bibnamefont {Babbush}},
  \bibinfo {author} {\bibfnamefont {Vadim}\ \bibnamefont {Smelyanskiy}},
  \bibinfo {author} {\bibfnamefont {John}\ \bibnamefont {Martinis}}, \ and\
  \bibinfo {author} {\bibfnamefont {Hartmut}\ \bibnamefont {Neven}},\
  }\bibfield  {title} {\enquote {\bibinfo {title} {What is the computational
  value of finite-range tunneling?}}\ }\href {\doibase
  10.1103/PhysRevX.6.031015} {\bibfield  {journal} {\bibinfo  {journal} {Phys.
  Rev. X}\ }\textbf {\bibinfo {volume} {6}},\ \bibinfo {pages} {031015}
  (\bibinfo {year} {2016})}\BibitemShut {NoStop}%
\bibitem [{\citenamefont {Farhi}\ \emph
  {et~al.}(2014{\natexlab{a}})\citenamefont {Farhi}, \citenamefont
  {Goldstone},\ and\ \citenamefont {Gutmann}}]{qaoa}%
  \BibitemOpen
  \bibfield  {author} {\bibinfo {author} {\bibfnamefont {Edward}\ \bibnamefont
  {Farhi}}, \bibinfo {author} {\bibfnamefont {Jeffrey}\ \bibnamefont
  {Goldstone}}, \ and\ \bibinfo {author} {\bibfnamefont {Sam}\ \bibnamefont
  {Gutmann}},\ }\bibfield  {title} {\enquote {\bibinfo {title} {A quantum
  approximate optimization algorithm},}\ }\href@noop {} {\  (\bibinfo {year}
  {2014}{\natexlab{a}})},\ \Eprint {http://arxiv.org/abs/1411.4028}
  {arXiv:1411.4028 [quant-ph]} \BibitemShut {NoStop}%
\bibitem [{\citenamefont {Farhi}\ \emph
  {et~al.}(2014{\natexlab{b}})\citenamefont {Farhi}, \citenamefont
  {Goldstone},\ and\ \citenamefont {Gutmann}}]{qaoa2}%
  \BibitemOpen
  \bibfield  {author} {\bibinfo {author} {\bibfnamefont {Edward}\ \bibnamefont
  {Farhi}}, \bibinfo {author} {\bibfnamefont {Jeffrey}\ \bibnamefont
  {Goldstone}}, \ and\ \bibinfo {author} {\bibfnamefont {Sam}\ \bibnamefont
  {Gutmann}},\ }\bibfield  {title} {\enquote {\bibinfo {title} {A quantum
  approximate optimization algorithm applied to a bounded occurrence constraint
  problem},}\ }\href@noop {} {\  (\bibinfo {year} {2014}{\natexlab{b}})},\
  \Eprint {http://arxiv.org/abs/1412.6062} {arXiv:1412.6062 [quant-ph]}
  \BibitemShut {NoStop}%
\bibitem [{\citenamefont {Biswas}\ \emph {et~al.}(2017)\citenamefont {Biswas},
  \citenamefont {Jiang}, \citenamefont {Kechezhi}, \citenamefont {Knysh},
  \citenamefont {Mandr{\`a}}, \citenamefont {O{'}Gorman}, \citenamefont
  {Perdomo-Ortiz}, \citenamefont {Petukhov}, \citenamefont {Realpe-G{\'o}mez},
  \citenamefont {Rieffel}, \citenamefont {Venturelli}, \citenamefont {Vasko},\
  and\ \citenamefont {Wang}}]{nasa-qc-perspective}%
  \BibitemOpen
  \bibfield  {author} {\bibinfo {author} {\bibfnamefont {Rupak}\ \bibnamefont
  {Biswas}}, \bibinfo {author} {\bibfnamefont {Zhang}\ \bibnamefont {Jiang}},
  \bibinfo {author} {\bibfnamefont {Kostya}\ \bibnamefont {Kechezhi}}, \bibinfo
  {author} {\bibfnamefont {Sergey}\ \bibnamefont {Knysh}}, \bibinfo {author}
  {\bibfnamefont {Salvatore}\ \bibnamefont {Mandr{\`a}}}, \bibinfo {author}
  {\bibfnamefont {Bryan}\ \bibnamefont {O{'}Gorman}}, \bibinfo {author}
  {\bibfnamefont {Alejandro}\ \bibnamefont {Perdomo-Ortiz}}, \bibinfo {author}
  {\bibfnamefont {Andre}\ \bibnamefont {Petukhov}}, \bibinfo {author}
  {\bibfnamefont {John}\ \bibnamefont {Realpe-G{\'o}mez}}, \bibinfo {author}
  {\bibfnamefont {Eleanor}\ \bibnamefont {Rieffel}}, \bibinfo {author}
  {\bibfnamefont {Davide}\ \bibnamefont {Venturelli}}, \bibinfo {author}
  {\bibfnamefont {Fedir}\ \bibnamefont {Vasko}}, \ and\ \bibinfo {author}
  {\bibfnamefont {Zhihui}\ \bibnamefont {Wang}},\ }\bibfield  {title} {\enquote
  {\bibinfo {title} {A {NASA} perspective on quantum computing: Opportunities
  and challenges},}\ }\href {\doibase 10.1016/j.parco.2016.11.002} {\bibfield
  {journal} {\bibinfo  {journal} {Parallel Computing}\ }\textbf {\bibinfo
  {volume} {64}},\ \bibinfo {pages} {81--98} (\bibinfo {year}
  {2017})}\BibitemShut {NoStop}%
\bibitem [{\citenamefont {Wecker}\ \emph {et~al.}(2016)\citenamefont {Wecker},
  \citenamefont {Hastings},\ and\ \citenamefont
  {Troyer}}]{wecker_training_2016}%
  \BibitemOpen
  \bibfield  {author} {\bibinfo {author} {\bibfnamefont {Dave}\ \bibnamefont
  {Wecker}}, \bibinfo {author} {\bibfnamefont {Matthew~B.}\ \bibnamefont
  {Hastings}}, \ and\ \bibinfo {author} {\bibfnamefont {Matthias}\ \bibnamefont
  {Troyer}},\ }\bibfield  {title} {\enquote {\bibinfo {title} {Training a
  quantum optimizer},}\ }\href {\doibase 10.1103/PhysRevA.94.022309} {\bibfield
   {journal} {\bibinfo  {journal} {Phys. Rev. A}\ }\textbf {\bibinfo {volume}
  {94}},\ \bibinfo {pages} {022309} (\bibinfo {year} {2016})}\BibitemShut
  {NoStop}%
\bibitem [{\citenamefont {Farhi}\ and\ \citenamefont
  {Harrow}(2016)}]{qaoa-supremacy}%
  \BibitemOpen
  \bibfield  {author} {\bibinfo {author} {\bibfnamefont {Edward}\ \bibnamefont
  {Farhi}}\ and\ \bibinfo {author} {\bibfnamefont {Aram~W}\ \bibnamefont
  {Harrow}},\ }\bibfield  {title} {\enquote {\bibinfo {title} {Quantum
  supremacy through the quantum approximate optimization algorithm},}\
  }\href@noop {} {\  (\bibinfo {year} {2016})},\ \Eprint
  {http://arxiv.org/abs/1602.07674} {arXiv:1602.07674 [quant-ph]} \BibitemShut
  {NoStop}%
\bibitem [{\citenamefont {Jiang}\ \emph {et~al.}(2017)\citenamefont {Jiang},
  \citenamefont {Rieffel},\ and\ \citenamefont {Wang}}]{qaoa-grovers}%
  \BibitemOpen
  \bibfield  {author} {\bibinfo {author} {\bibfnamefont {Zhang}\ \bibnamefont
  {Jiang}}, \bibinfo {author} {\bibfnamefont {Eleanor~G.}\ \bibnamefont
  {Rieffel}}, \ and\ \bibinfo {author} {\bibfnamefont {Zhihui}\ \bibnamefont
  {Wang}},\ }\bibfield  {title} {\enquote {\bibinfo {title} {Near-optimal
  quantum circuit for {Grover's} unstructured search using a transverse
  field},}\ }\href {\doibase 10.1103/PhysRevA.95.062317} {\bibfield  {journal}
  {\bibinfo  {journal} {Phys. Rev. A}\ }\textbf {\bibinfo {volume} {95}},\
  \bibinfo {pages} {062317} (\bibinfo {year} {2017})}\BibitemShut {NoStop}%
\bibitem [{\citenamefont {Wang}\ \emph {et~al.}(2018)\citenamefont {Wang},
  \citenamefont {Hadfield}, \citenamefont {Jiang},\ and\ \citenamefont
  {Rieffel}}]{qaoa-fermionic}%
  \BibitemOpen
  \bibfield  {author} {\bibinfo {author} {\bibfnamefont {Zhihui}\ \bibnamefont
  {Wang}}, \bibinfo {author} {\bibfnamefont {Stuart}\ \bibnamefont {Hadfield}},
  \bibinfo {author} {\bibfnamefont {Zhang}\ \bibnamefont {Jiang}}, \ and\
  \bibinfo {author} {\bibfnamefont {Eleanor~G.}\ \bibnamefont {Rieffel}},\
  }\bibfield  {title} {\enquote {\bibinfo {title} {Quantum approximate
  optimization algorithm for {MaxCut}: A fermionic view},}\ }\href {\doibase
  10.1103/PhysRevA.97.022304} {\bibfield  {journal} {\bibinfo  {journal} {Phys.
  Rev. A}\ }\textbf {\bibinfo {volume} {97}},\ \bibinfo {pages} {022304}
  (\bibinfo {year} {2018})}\BibitemShut {NoStop}%
\bibitem [{\citenamefont {Hadfield}\ \emph {et~al.}(2017)\citenamefont
  {Hadfield}, \citenamefont {Wang}, \citenamefont {O'Gorman}, \citenamefont
  {Rieffel}, \citenamefont {Venturelli},\ and\ \citenamefont
  {Biswas}}]{alternating-operator-ansatz}%
  \BibitemOpen
  \bibfield  {author} {\bibinfo {author} {\bibfnamefont {Stuart}\ \bibnamefont
  {Hadfield}}, \bibinfo {author} {\bibfnamefont {Zhihui}\ \bibnamefont {Wang}},
  \bibinfo {author} {\bibfnamefont {Bryan}\ \bibnamefont {O'Gorman}}, \bibinfo
  {author} {\bibfnamefont {Eleanor}\ \bibnamefont {Rieffel}}, \bibinfo {author}
  {\bibfnamefont {Davide}\ \bibnamefont {Venturelli}}, \ and\ \bibinfo {author}
  {\bibfnamefont {Rupak}\ \bibnamefont {Biswas}},\ }\bibfield  {title}
  {\enquote {\bibinfo {title} {From the quantum approximate optimization
  algorithm to a quantum alternating operator ansatz},}\ }\href {\doibase
  10.3390/a12020034} {\bibfield  {journal} {\bibinfo  {journal} {Algorithms}\
  }\textbf {\bibinfo {volume} {12}},\ \bibinfo {pages} {34} (\bibinfo {year}
  {2017})}\BibitemShut {NoStop}%
\bibitem [{\citenamefont {Lloyd}(2018)}]{qaoa-universal}%
  \BibitemOpen
  \bibfield  {author} {\bibinfo {author} {\bibfnamefont {Seth}\ \bibnamefont
  {Lloyd}},\ }\bibfield  {title} {\enquote {\bibinfo {title} {Quantum
  approximate optimization is computationally universal},}\ }\href@noop {} {\
  (\bibinfo {year} {2018})},\ \Eprint {http://arxiv.org/abs/1812.11075}
  {arXiv:1812.11075 [quant-ph]} \BibitemShut {NoStop}%
\bibitem [{\citenamefont {Farhi}\ \emph {et~al.}(2019)\citenamefont {Farhi},
  \citenamefont {Goldstone}, \citenamefont {Gutmann},\ and\ \citenamefont
  {Zhou}}]{farhi-sk-qaoa}%
  \BibitemOpen
  \bibfield  {author} {\bibinfo {author} {\bibfnamefont {Edward}\ \bibnamefont
  {Farhi}}, \bibinfo {author} {\bibfnamefont {Jeffrey}\ \bibnamefont
  {Goldstone}}, \bibinfo {author} {\bibfnamefont {Sam}\ \bibnamefont
  {Gutmann}}, \ and\ \bibinfo {author} {\bibfnamefont {Leo}\ \bibnamefont
  {Zhou}},\ }\bibfield  {title} {\enquote {\bibinfo {title} {The quantum
  approximate optimization algorithm and the {Sherrington-Kirkpatrick} model at
  infinite size},}\ }\href@noop {} {\  (\bibinfo {year} {2019})},\ \Eprint
  {http://arxiv.org/abs/1910.08187} {arXiv:1910.08187 [quant-ph]} \BibitemShut
  {NoStop}%
\bibitem [{\citenamefont {Otterbach}\ \emph {et~al.}(2017)\citenamefont
  {Otterbach}, \citenamefont {Manenti}, \citenamefont {Alidoust}, \citenamefont
  {Bestwick}, \citenamefont {Block}, \citenamefont {Bloom}, \citenamefont
  {Caldwell}, \citenamefont {Didier}, \citenamefont {Fried}, \citenamefont
  {Hong}, \citenamefont {Karalekas}, \citenamefont {Osborn}, \citenamefont
  {Papageorge}, \citenamefont {Peterson}, \citenamefont {Prawiroatmodjo},
  \citenamefont {Rubin}, \citenamefont {Ryan}, \citenamefont {Scarabelli},
  \citenamefont {Scheer}, \citenamefont {Sete}, \citenamefont {Sivarajah},
  \citenamefont {Smith}, \citenamefont {Staley}, \citenamefont {Tezak},
  \citenamefont {Zeng}, \citenamefont {Hudson}, \citenamefont {Johnson},
  \citenamefont {Reagor}, \citenamefont {Silva},\ and\ \citenamefont
  {Rigetti}}]{rigetti-qaoa-clustering}%
  \BibitemOpen
  \bibfield  {author} {\bibinfo {author} {\bibfnamefont {J.~S.}\ \bibnamefont
  {Otterbach}}, \bibinfo {author} {\bibfnamefont {R.}~\bibnamefont {Manenti}},
  \bibinfo {author} {\bibfnamefont {N.}~\bibnamefont {Alidoust}}, \bibinfo
  {author} {\bibfnamefont {A.}~\bibnamefont {Bestwick}}, \bibinfo {author}
  {\bibfnamefont {M.}~\bibnamefont {Block}}, \bibinfo {author} {\bibfnamefont
  {B.}~\bibnamefont {Bloom}}, \bibinfo {author} {\bibfnamefont
  {S.}~\bibnamefont {Caldwell}}, \bibinfo {author} {\bibfnamefont
  {N.}~\bibnamefont {Didier}}, \bibinfo {author} {\bibfnamefont {E.~Schuyler}\
  \bibnamefont {Fried}}, \bibinfo {author} {\bibfnamefont {S.}~\bibnamefont
  {Hong}}, \bibinfo {author} {\bibfnamefont {P.}~\bibnamefont {Karalekas}},
  \bibinfo {author} {\bibfnamefont {C.~B.}\ \bibnamefont {Osborn}}, \bibinfo
  {author} {\bibfnamefont {A.}~\bibnamefont {Papageorge}}, \bibinfo {author}
  {\bibfnamefont {E.~C.}\ \bibnamefont {Peterson}}, \bibinfo {author}
  {\bibfnamefont {G.}~\bibnamefont {Prawiroatmodjo}}, \bibinfo {author}
  {\bibfnamefont {N.}~\bibnamefont {Rubin}}, \bibinfo {author} {\bibfnamefont
  {Colm~A.}\ \bibnamefont {Ryan}}, \bibinfo {author} {\bibfnamefont
  {D.}~\bibnamefont {Scarabelli}}, \bibinfo {author} {\bibfnamefont
  {M.}~\bibnamefont {Scheer}}, \bibinfo {author} {\bibfnamefont {E.~A.}\
  \bibnamefont {Sete}}, \bibinfo {author} {\bibfnamefont {P.}~\bibnamefont
  {Sivarajah}}, \bibinfo {author} {\bibfnamefont {Robert~S.}\ \bibnamefont
  {Smith}}, \bibinfo {author} {\bibfnamefont {A.}~\bibnamefont {Staley}},
  \bibinfo {author} {\bibfnamefont {N.}~\bibnamefont {Tezak}}, \bibinfo
  {author} {\bibfnamefont {W.~J.}\ \bibnamefont {Zeng}}, \bibinfo {author}
  {\bibfnamefont {A.}~\bibnamefont {Hudson}}, \bibinfo {author} {\bibfnamefont
  {Blake~R.}\ \bibnamefont {Johnson}}, \bibinfo {author} {\bibfnamefont
  {M.}~\bibnamefont {Reagor}}, \bibinfo {author} {\bibfnamefont {M.~P.~da}\
  \bibnamefont {Silva}}, \ and\ \bibinfo {author} {\bibfnamefont
  {C.}~\bibnamefont {Rigetti}},\ }\bibfield  {title} {\enquote {\bibinfo
  {title} {Unsupervised machine learning on a hybrid quantum computer},}\
  }\href@noop {} {\  (\bibinfo {year} {2017})},\ \Eprint
  {http://arxiv.org/abs/1712.05771} {arXiv:1712.05771 [quant-ph]} \BibitemShut
  {NoStop}%
\bibitem [{\citenamefont {Willsch}\ \emph {et~al.}(2019)\citenamefont
  {Willsch}, \citenamefont {Willsch}, \citenamefont {Jin}, \citenamefont
  {Raedt},\ and\ \citenamefont {Michielsen}}]{2019-julich-qaoa-on-ibm}%
  \BibitemOpen
  \bibfield  {author} {\bibinfo {author} {\bibfnamefont {Madita}\ \bibnamefont
  {Willsch}}, \bibinfo {author} {\bibfnamefont {Dennis}\ \bibnamefont
  {Willsch}}, \bibinfo {author} {\bibfnamefont {Fengping}\ \bibnamefont {Jin}},
  \bibinfo {author} {\bibfnamefont {Hans~De}\ \bibnamefont {Raedt}}, \ and\
  \bibinfo {author} {\bibfnamefont {Kristel}\ \bibnamefont {Michielsen}},\
  }\bibfield  {title} {\enquote {\bibinfo {title} {Benchmarking the quantum
  approximate optimization algorithm},}\ }\href@noop {} {\  (\bibinfo {year}
  {2019})},\ \Eprint {http://arxiv.org/abs/1907.02359} {arXiv:1907.02359
  [quant-ph]} \BibitemShut {NoStop}%
\bibitem [{\citenamefont {Abrams}\ \emph {et~al.}(2019)\citenamefont {Abrams},
  \citenamefont {Didier}, \citenamefont {Johnson}, \citenamefont {Silva},\ and\
  \citenamefont {Ryan}}]{rigetti-xy-gate}%
  \BibitemOpen
  \bibfield  {author} {\bibinfo {author} {\bibfnamefont {Deanna~M.}\
  \bibnamefont {Abrams}}, \bibinfo {author} {\bibfnamefont {Nicolas}\
  \bibnamefont {Didier}}, \bibinfo {author} {\bibfnamefont {Blake~R.}\
  \bibnamefont {Johnson}}, \bibinfo {author} {\bibfnamefont {Marcus P.~da}\
  \bibnamefont {Silva}}, \ and\ \bibinfo {author} {\bibfnamefont {Colm~A.}\
  \bibnamefont {Ryan}},\ }\bibfield  {title} {\enquote {\bibinfo {title}
  {Implementation of the {XY} interaction family with calibration of a single
  pulse},}\ }\href@noop {} {\  (\bibinfo {year} {2019})},\ \Eprint
  {http://arxiv.org/abs/1912.04424} {arXiv:1912.04424 [quant-ph]} \BibitemShut
  {NoStop}%
\bibitem [{\citenamefont {Bengtsson}\ \emph {et~al.}(2019)\citenamefont
  {Bengtsson}, \citenamefont {Vikst{{\aa}}l}, \citenamefont {Warren},
  \citenamefont {Svensson}, \citenamefont {Gu}, \citenamefont {Kockum},
  \citenamefont {Krantz}, \citenamefont {Križan}, \citenamefont {Shiri},
  \citenamefont {Svensson}, \citenamefont {Tancredi}, \citenamefont
  {Johansson}, \citenamefont {Delsing}, \citenamefont {Ferrini},\ and\
  \citenamefont {Bylander}}]{chalmers-2q-qaoa}%
  \BibitemOpen
  \bibfield  {author} {\bibinfo {author} {\bibfnamefont {Andreas}\ \bibnamefont
  {Bengtsson}}, \bibinfo {author} {\bibfnamefont {Pontus}\ \bibnamefont
  {Vikst{{\aa}}l}}, \bibinfo {author} {\bibfnamefont {Christopher}\
  \bibnamefont {Warren}}, \bibinfo {author} {\bibfnamefont {Marika}\
  \bibnamefont {Svensson}}, \bibinfo {author} {\bibfnamefont {Xiu}\
  \bibnamefont {Gu}}, \bibinfo {author} {\bibfnamefont {Anton~Frisk}\
  \bibnamefont {Kockum}}, \bibinfo {author} {\bibfnamefont {Philip}\
  \bibnamefont {Krantz}}, \bibinfo {author} {\bibfnamefont {Christian}\
  \bibnamefont {Križan}}, \bibinfo {author} {\bibfnamefont {Daryoush}\
  \bibnamefont {Shiri}}, \bibinfo {author} {\bibfnamefont {Ida-Maria}\
  \bibnamefont {Svensson}}, \bibinfo {author} {\bibfnamefont {Giovanna}\
  \bibnamefont {Tancredi}}, \bibinfo {author} {\bibfnamefont {G{\"o}ran}\
  \bibnamefont {Johansson}}, \bibinfo {author} {\bibfnamefont {Per}\
  \bibnamefont {Delsing}}, \bibinfo {author} {\bibfnamefont {Giulia}\
  \bibnamefont {Ferrini}}, \ and\ \bibinfo {author} {\bibfnamefont {Jonas}\
  \bibnamefont {Bylander}},\ }\bibfield  {title} {\enquote {\bibinfo {title}
  {Quantum approximate optimization of the exact-cover problem on a
  superconducting quantum processor},}\ }\href@noop {} {\  (\bibinfo {year}
  {2019})},\ \Eprint {http://arxiv.org/abs/1912.10495} {arXiv:1912.10495
  [quant-ph]} \BibitemShut {NoStop}%
\bibitem [{\citenamefont {Pagano}\ \emph {et~al.}(2019)\citenamefont {Pagano},
  \citenamefont {Bapat}, \citenamefont {Becker}, \citenamefont {Collins},
  \citenamefont {De}, \citenamefont {Hess}, \citenamefont {Kaplan},
  \citenamefont {Kyprianidis}, \citenamefont {Tan}, \citenamefont {Baldwin},
  \citenamefont {Brady}, \citenamefont {Deshpande}, \citenamefont {Liu},
  \citenamefont {Jordan}, \citenamefont {Gorshkov},\ and\ \citenamefont
  {Monroe}}]{2019-monroe-qaoa}%
  \BibitemOpen
  \bibfield  {author} {\bibinfo {author} {\bibfnamefont {G.}~\bibnamefont
  {Pagano}}, \bibinfo {author} {\bibfnamefont {A.}~\bibnamefont {Bapat}},
  \bibinfo {author} {\bibfnamefont {P.}~\bibnamefont {Becker}}, \bibinfo
  {author} {\bibfnamefont {K.~S.}\ \bibnamefont {Collins}}, \bibinfo {author}
  {\bibfnamefont {A.}~\bibnamefont {De}}, \bibinfo {author} {\bibfnamefont
  {P.~W.}\ \bibnamefont {Hess}}, \bibinfo {author} {\bibfnamefont {H.~B.}\
  \bibnamefont {Kaplan}}, \bibinfo {author} {\bibfnamefont {A.}~\bibnamefont
  {Kyprianidis}}, \bibinfo {author} {\bibfnamefont {W.~L.}\ \bibnamefont
  {Tan}}, \bibinfo {author} {\bibfnamefont {C.}~\bibnamefont {Baldwin}},
  \bibinfo {author} {\bibfnamefont {L.~T.}\ \bibnamefont {Brady}}, \bibinfo
  {author} {\bibfnamefont {A.}~\bibnamefont {Deshpande}}, \bibinfo {author}
  {\bibfnamefont {F.}~\bibnamefont {Liu}}, \bibinfo {author} {\bibfnamefont
  {S.}~\bibnamefont {Jordan}}, \bibinfo {author} {\bibfnamefont {A.~V.}\
  \bibnamefont {Gorshkov}}, \ and\ \bibinfo {author} {\bibfnamefont
  {C.}~\bibnamefont {Monroe}},\ }\bibfield  {title} {\enquote {\bibinfo {title}
  {Quantum approximate optimization with a trapped-ion quantum simulator},}\
  }\href@noop {} {\  (\bibinfo {year} {2019})},\ \Eprint
  {http://arxiv.org/abs/1906.02700} {arXiv:1906.02700 [quant-ph]} \BibitemShut
  {NoStop}%
\bibitem [{\citenamefont {Qiang}\ \emph {et~al.}(2018)\citenamefont {Qiang},
  \citenamefont {Zhou}, \citenamefont {Wang}, \citenamefont {Wilkes},
  \citenamefont {Loke}, \citenamefont {O{'}Gara}, \citenamefont {Kling},
  \citenamefont {Marshall}, \citenamefont {Santagati}, \citenamefont {Ralph},
  \citenamefont {Wang}, \citenamefont {O{'}Brien}, \citenamefont {Thompson},\
  and\ \citenamefont {Matthews}}]{2018-bristol-photonics}%
  \BibitemOpen
  \bibfield  {author} {\bibinfo {author} {\bibfnamefont {Xiaogang}\
  \bibnamefont {Qiang}}, \bibinfo {author} {\bibfnamefont {Xiaoqi}\
  \bibnamefont {Zhou}}, \bibinfo {author} {\bibfnamefont {Jianwei}\
  \bibnamefont {Wang}}, \bibinfo {author} {\bibfnamefont {Callum~M.}\
  \bibnamefont {Wilkes}}, \bibinfo {author} {\bibfnamefont {Thomas}\
  \bibnamefont {Loke}}, \bibinfo {author} {\bibfnamefont {Sean}\ \bibnamefont
  {O{'}Gara}}, \bibinfo {author} {\bibfnamefont {Laurent}\ \bibnamefont
  {Kling}}, \bibinfo {author} {\bibfnamefont {Graham~D.}\ \bibnamefont
  {Marshall}}, \bibinfo {author} {\bibfnamefont {Raffaele}\ \bibnamefont
  {Santagati}}, \bibinfo {author} {\bibfnamefont {Timothy~C.}\ \bibnamefont
  {Ralph}}, \bibinfo {author} {\bibfnamefont {Jingbo~B.}\ \bibnamefont {Wang}},
  \bibinfo {author} {\bibfnamefont {Jeremy~L.}\ \bibnamefont {O{'}Brien}},
  \bibinfo {author} {\bibfnamefont {Mark~G.}\ \bibnamefont {Thompson}}, \ and\
  \bibinfo {author} {\bibfnamefont {Jonathan C.~F.}\ \bibnamefont {Matthews}},\
  }\bibfield  {title} {\enquote {\bibinfo {title} {Large-scale silicon quantum
  photonics implementing arbitrary two-qubit processing},}\ }\href {\doibase
  10.1038/s41566-018-0236-y} {\bibfield  {journal} {\bibinfo  {journal} {Nature
  Photonics}\ }\textbf {\bibinfo {volume} {12}},\ \bibinfo {pages} {534--539}
  (\bibinfo {year} {2018})}\BibitemShut {NoStop}%
\bibitem [{\citenamefont {{The Cirq Developers}}(2020)}]{cirq}%
  \BibitemOpen
  \bibfield  {author} {\bibinfo {author} {\bibnamefont {{The Cirq
  Developers}}},\ }\href {https://github.com/quantumlib/Cirq} {\enquote
  {\bibinfo {title} {Cirq: A python framework for creating, editing, and
  invoking noisy intermediate scale quantum ({NISQ}) circuits},}\ } (\bibinfo
  {year} {2020}),\ \bibinfo {note}
  {https://github.com/quantumlib/Cirq}\BibitemShut {NoStop}%
\bibitem [{\citenamefont {R{\o}nnow}\ \emph {et~al.}(2014)\citenamefont
  {R{\o}nnow}, \citenamefont {Wang}, \citenamefont {Job}, \citenamefont
  {Boixo}, \citenamefont {Isakov}, \citenamefont {Wecker}, \citenamefont
  {Martinis}, \citenamefont {Lidar},\ and\ \citenamefont
  {Troyer}}]{ronnow_defining_2014}%
  \BibitemOpen
  \bibfield  {author} {\bibinfo {author} {\bibfnamefont {Troels~F.}\
  \bibnamefont {R{\o}nnow}}, \bibinfo {author} {\bibfnamefont {Zhihui}\
  \bibnamefont {Wang}}, \bibinfo {author} {\bibfnamefont {Joshua}\ \bibnamefont
  {Job}}, \bibinfo {author} {\bibfnamefont {Sergio}\ \bibnamefont {Boixo}},
  \bibinfo {author} {\bibfnamefont {Sergei~V.}\ \bibnamefont {Isakov}},
  \bibinfo {author} {\bibfnamefont {David}\ \bibnamefont {Wecker}}, \bibinfo
  {author} {\bibfnamefont {John~M.}\ \bibnamefont {Martinis}}, \bibinfo
  {author} {\bibfnamefont {Daniel~A.}\ \bibnamefont {Lidar}}, \ and\ \bibinfo
  {author} {\bibfnamefont {Matthias}\ \bibnamefont {Troyer}},\ }\bibfield
  {title} {\enquote {\bibinfo {title} {Defining and detecting quantum
  speedup},}\ }\href {\doibase 10.1126/science.1252319} {\bibfield  {journal}
  {\bibinfo  {journal} {Science}\ }\textbf {\bibinfo {volume} {345}},\ \bibinfo
  {pages} {420--424} (\bibinfo {year} {2014})}\BibitemShut {NoStop}%
\bibitem [{\citenamefont {Sherrington}\ and\ \citenamefont
  {Kirkpatrick}(1975)}]{sherrington_kirkpatrick_model_1975}%
  \BibitemOpen
  \bibfield  {author} {\bibinfo {author} {\bibfnamefont {David}\ \bibnamefont
  {Sherrington}}\ and\ \bibinfo {author} {\bibfnamefont {Scott}\ \bibnamefont
  {Kirkpatrick}},\ }\bibfield  {title} {\enquote {\bibinfo {title} {Solvable
  model of a spin-glass},}\ }\href {\doibase 10.1103/PhysRevLett.35.1792}
  {\bibfield  {journal} {\bibinfo  {journal} {Phys. Rev. Lett.}\ }\textbf
  {\bibinfo {volume} {35}},\ \bibinfo {pages} {1792--1796} (\bibinfo {year}
  {1975})}\BibitemShut {NoStop}%
\bibitem [{\citenamefont {Hirata}\ \emph {et~al.}(2011)\citenamefont {Hirata},
  \citenamefont {Nakanishi}, \citenamefont {Yamashita},\ and\ \citenamefont
  {Nakashima}}]{hirata2011efficient}%
  \BibitemOpen
  \bibfield  {author} {\bibinfo {author} {\bibfnamefont {Yuichi}\ \bibnamefont
  {Hirata}}, \bibinfo {author} {\bibfnamefont {Masaki}\ \bibnamefont
  {Nakanishi}}, \bibinfo {author} {\bibfnamefont {Shigeru}\ \bibnamefont
  {Yamashita}}, \ and\ \bibinfo {author} {\bibfnamefont {Yasuhiko}\
  \bibnamefont {Nakashima}},\ }\bibfield  {title} {\enquote {\bibinfo {title}
  {An efficient conversion of quantum circuits to a linear nearest neighbor
  architecture},}\ }\href {https://dl.acm.org/doi/10.5555/2011383.2011393}
  {\bibfield  {journal} {\bibinfo  {journal} {Quantum Info. Comput.}\ }\textbf
  {\bibinfo {volume} {11}},\ \bibinfo {pages} {142–166} (\bibinfo {year}
  {2011})}\BibitemShut {NoStop}%
\bibitem [{\citenamefont {Goemans}\ and\ \citenamefont
  {Williamson}(1995)}]{gw-maxcut}%
  \BibitemOpen
  \bibfield  {author} {\bibinfo {author} {\bibfnamefont {Michel~X.}\
  \bibnamefont {Goemans}}\ and\ \bibinfo {author} {\bibfnamefont {David~P.}\
  \bibnamefont {Williamson}},\ }\bibfield  {title} {\enquote {\bibinfo {title}
  {Improved approximation algorithms for maximum cut and satisfiability
  problems using semidefinite programming},}\ }\href {\doibase
  10.1145/227683.227684} {\bibfield  {journal} {\bibinfo  {journal} {J. ACM}\
  }\textbf {\bibinfo {volume} {42}},\ \bibinfo {pages} {1115–1145} (\bibinfo
  {year} {1995})}\BibitemShut {NoStop}%
\bibitem [{\citenamefont {Bravyi}\ \emph {et~al.}(2019)\citenamefont {Bravyi},
  \citenamefont {Kliesch}, \citenamefont {Koenig},\ and\ \citenamefont
  {Tang}}]{2019-bravyi-qaoa}%
  \BibitemOpen
  \bibfield  {author} {\bibinfo {author} {\bibfnamefont {Sergey}\ \bibnamefont
  {Bravyi}}, \bibinfo {author} {\bibfnamefont {Alexander}\ \bibnamefont
  {Kliesch}}, \bibinfo {author} {\bibfnamefont {Robert}\ \bibnamefont
  {Koenig}}, \ and\ \bibinfo {author} {\bibfnamefont {Eugene}\ \bibnamefont
  {Tang}},\ }\bibfield  {title} {\enquote {\bibinfo {title} {Obstacles to state
  preparation and variational optimization from symmetry protection},}\
  }\href@noop {} {\  (\bibinfo {year} {2019})},\ \Eprint
  {http://arxiv.org/abs/1910.08980} {arXiv:1910.08980 [quant-ph]} \BibitemShut
  {NoStop}%
\bibitem [{\citenamefont {Cowtan}\ \emph {et~al.}(2019)\citenamefont {Cowtan},
  \citenamefont {Dilkes}, \citenamefont {Duncan}, \citenamefont {Krajenbrink},
  \citenamefont {Simmons},\ and\ \citenamefont
  {Sivarajah}}]{tket-qubit-routing}%
  \BibitemOpen
  \bibfield  {author} {\bibinfo {author} {\bibfnamefont {Alexander}\
  \bibnamefont {Cowtan}}, \bibinfo {author} {\bibfnamefont {Silas}\
  \bibnamefont {Dilkes}}, \bibinfo {author} {\bibfnamefont {Ross}\ \bibnamefont
  {Duncan}}, \bibinfo {author} {\bibfnamefont {Alexandre}\ \bibnamefont
  {Krajenbrink}}, \bibinfo {author} {\bibfnamefont {Will}\ \bibnamefont
  {Simmons}}, \ and\ \bibinfo {author} {\bibfnamefont {Seyon}\ \bibnamefont
  {Sivarajah}},\ }\bibfield  {title} {\enquote {\bibinfo {title} {On the qubit
  routing problem},}\ }\href@noop {} {\  (\bibinfo {year} {2019})},\ \Eprint
  {http://arxiv.org/abs/1902.08091} {arXiv:1902.08091 [quant-ph]} \BibitemShut
  {NoStop}%
\bibitem [{\citenamefont {Yang}\ \emph {et~al.}(2017)\citenamefont {Yang},
  \citenamefont {Rahmani}, \citenamefont {Shabani}, \citenamefont {Neven},\
  and\ \citenamefont {Chamon}}]{pontryagin-angle-optimization}%
  \BibitemOpen
  \bibfield  {author} {\bibinfo {author} {\bibfnamefont {Zhi-Cheng}\
  \bibnamefont {Yang}}, \bibinfo {author} {\bibfnamefont {Armin}\ \bibnamefont
  {Rahmani}}, \bibinfo {author} {\bibfnamefont {Alireza}\ \bibnamefont
  {Shabani}}, \bibinfo {author} {\bibfnamefont {Hartmut}\ \bibnamefont
  {Neven}}, \ and\ \bibinfo {author} {\bibfnamefont {Claudio}\ \bibnamefont
  {Chamon}},\ }\bibfield  {title} {\enquote {\bibinfo {title} {Optimizing
  variational quantum algorithms using {Pontryagin}'s minimum principle},}\
  }\href {\doibase 10.1103/PhysRevX.7.021027} {\bibfield  {journal} {\bibinfo
  {journal} {Phys. Rev. X}\ }\textbf {\bibinfo {volume} {7}},\ \bibinfo {pages}
  {021027} (\bibinfo {year} {2017})}\BibitemShut {NoStop}%
\bibitem [{\citenamefont {McClean}\ \emph {et~al.}(2018)\citenamefont
  {McClean}, \citenamefont {Boixo}, \citenamefont {Smelyanskiy}, \citenamefont
  {Babbush},\ and\ \citenamefont {Neven}}]{barren-plateaus}%
  \BibitemOpen
  \bibfield  {author} {\bibinfo {author} {\bibfnamefont {Jarrod~R.}\
  \bibnamefont {McClean}}, \bibinfo {author} {\bibfnamefont {Sergio}\
  \bibnamefont {Boixo}}, \bibinfo {author} {\bibfnamefont {Vadim~N.}\
  \bibnamefont {Smelyanskiy}}, \bibinfo {author} {\bibfnamefont {Ryan}\
  \bibnamefont {Babbush}}, \ and\ \bibinfo {author} {\bibfnamefont {Hartmut}\
  \bibnamefont {Neven}},\ }\bibfield  {title} {\enquote {\bibinfo {title}
  {Barren plateaus in quantum neural network training landscapes},}\ }\href
  {\doibase 10.1038/s41467-018-07090-4} {\bibfield  {journal} {\bibinfo
  {journal} {Nature Communications}\ }\textbf {\bibinfo {volume} {9}},\
  \bibinfo {pages} {4812} (\bibinfo {year} {2018})}\BibitemShut {NoStop}%
\bibitem [{\citenamefont {Sung}\ \emph {et~al.}()\citenamefont {Sung},
  \citenamefont {Harrigan}, \citenamefont {Rubin}, \citenamefont {Jiang},
  \citenamefont {Babbush},\ and\ \citenamefont {McClean}}]{sung-optimizers}%
  \BibitemOpen
  \bibfield  {author} {\bibinfo {author} {\bibfnamefont {Kevin~J.}\
  \bibnamefont {Sung}}, \bibinfo {author} {\bibfnamefont {Matthew~P.}\
  \bibnamefont {Harrigan}}, \bibinfo {author} {\bibfnamefont {Nicholas}\
  \bibnamefont {Rubin}}, \bibinfo {author} {\bibfnamefont {Zhang}\ \bibnamefont
  {Jiang}}, \bibinfo {author} {\bibfnamefont {Ryan}\ \bibnamefont {Babbush}}, \
  and\ \bibinfo {author} {\bibfnamefont {Jarrod}\ \bibnamefont {McClean}},\
  }\bibfield  {title} {\enquote {\bibinfo {title} {Practical optimizers for
  variational quantum algorithms},}\ }\href@noop {} {\bibinfo  {journal} {in
  preparation}\ }\BibitemShut {NoStop}%
\bibitem [{\citenamefont {Quantum}\ and\ \citenamefont
  {Collaborators}(2020)}]{qaoadata}%
  \BibitemOpen
\bibfield  {journal} {  }\bibfield  {author} {\bibinfo {author} {\bibfnamefont
  {Google~AI}\ \bibnamefont {Quantum}}\ and\ \bibinfo {author} {\bibnamefont
  {Collaborators}},\ }\bibfield  {title} {\enquote {\bibinfo {title} {{Sycamore
  QAOA experimental data}},}\ }\href {\doibase 10.6084/m9.figshare.12597590.v2}
  {\  (\bibinfo {year} {2020}),\ 10.6084/m9.figshare.12597590.v2}\BibitemShut
  {NoStop}%
\bibitem [{\citenamefont {Zhang}\ \emph {et~al.}(2003)\citenamefont {Zhang},
  \citenamefont {Vala}, \citenamefont {Sastry},\ and\ \citenamefont
  {Whaley}}]{PhysRevA.67.042313}%
  \BibitemOpen
  \bibfield  {author} {\bibinfo {author} {\bibfnamefont {Jun}\ \bibnamefont
  {Zhang}}, \bibinfo {author} {\bibfnamefont {Jiri}\ \bibnamefont {Vala}},
  \bibinfo {author} {\bibfnamefont {Shankar}\ \bibnamefont {Sastry}}, \ and\
  \bibinfo {author} {\bibfnamefont {K.~Birgitta}\ \bibnamefont {Whaley}},\
  }\bibfield  {title} {\enquote {\bibinfo {title} {Geometric theory of nonlocal
  two-qubit operations},}\ }\href {\doibase 10.1103/PhysRevA.67.042313}
  {\bibfield  {journal} {\bibinfo  {journal} {Phys. Rev. A}\ }\textbf {\bibinfo
  {volume} {67}},\ \bibinfo {pages} {042313} (\bibinfo {year}
  {2003})}\BibitemShut {NoStop}%
\end{thebibliography}%


%
\onecolumngrid

\newpage
\appendix
\setcounter{table}{0}
\renewcommand{\thetable}{S\arabic{table}}
\setcounter{figure}{0}
\renewcommand{\thefigure}{S\arabic{figure}}

\section{Hardware and Compilation Details} \label{app:hardware-and-compilation}

In this section, we discuss detailed compilation of the desired unitaries into the hardware native
gateset, particularly the $\syc$ gate defined in \figa{circuit}{c}.
The $\syc$ gate is similar to the gate used in \citet{google_supremacy_2019} but with the conditional phase tuned to be precisely $\pi/6$.
A $\sqrtiswap$ gate is simultaneously calibrated and available but has a longer gate duration and requires additional (physical) Z rotations to match phases.
The required interactions for this study are compiled to an equivalent number of $\syc$ and $\sqrtiswap$, so $\syc$ was used in all circuits. Single-qubit microwave pulses enact ``Phased X'' gates $\PhX(\theta, \phi)$ (alternatively called XY rotations or the W gate) with $\phi=0$ corresponding to $R_X(\theta)$ and $\phi=\frac{\pi}{2}$ corresponding to $R_Y(\theta)$ (up to global phase). Intermediate values of $\phi$ control the axis of rotation in the X-Y plane of the Bloch sphere.

Arbitrary single-qubit rotations can be applied by a $\PhX(\theta, \phi)$ gate followed by a $R_Z(\vartheta)$ gate. As a compilation step, we merge adjacent single-qubit operations to be of this form. Therefore, our circuit is structured as a repeating sequence of: a layer of $\PhX$ gates; a layer of $Z$ gates; and a layer of $\syc$ gates. All Z rotations of the form $\exp \left[ -i \theta Z \right]$ can be efficiently commuted through $\syc$ and $\PhX$ to the end of the circuit and discarded. This leaves alternating layers of $\PhX$ and $\syc$ gates. The overheads of compilation are summarized in \tab{problems}.

\begin{table}[htbp!]
    \centering
    \begin{tabular}{p{7em} p{7em} p{7em} p{4em}}
        \hline
        Problem         &  Routing      & Interaction                       & Synthesis \\
        \hline\\[-1em]
        Hardware Grid   &  WESN         & $e^{-i \gamma ZZ}$                & 2         \\
        MaxCut          &  Greedy       & $e^{-i \gamma ZZ}$      & 2      \\
        MaxCut          &  Greedy       & $\swap$       & 3      \\
        SK Model        &  Swap Network & $e^{-i \gamma ZZ}\! \cdot \swap$  & 3         \\
        \hline
    \end{tabular}
    \caption{Compilation details for the problems studied. ``Routing'' gives the strategy used for routing, ``Interaction'' gives the type of two-qubit gates which need to be compiled, and ``Synthesis'' gives the number of hardware native 2-qubit $\syc$ gates required to realize the target interaction. ``WESN" routing refers to planar activation of West, East, etc. links.}
    \label{tab:problems}
\end{table}

\textbf{Compilation of $ZZ(\gamma)$}.
These interactions (used for Hardware Grid and MaxCut problems) can be compiled with 2 layers of $\syc$ gates and 2+1 associated layers of single qubit $\PhX$ gates.  We report the required number of single-qubit layers as 2+1 because the initial (or final) layer from one set of interactions
can be merged into the final (initial) single qubit gate layer of the preceding (following) set of interactions. In general, the number of single qubit layers will be equivalent to the number of two-qubit gate layers with one additional single-qubit layer at the beginning of the circuit and one additional single-qubit layer at the end of the circuit.
The explicit compilation of $ZZ$ to $\syc$ is available in Cirq and a proof can be found in the supplemental material of Ref.~\cite{google_supremacy_2019}. Here we reproduce the derivation in slightly different notation but following a similar motivation.

\begin{figure}[htbp!]
    \includegraphics[width=0.6\textwidth]{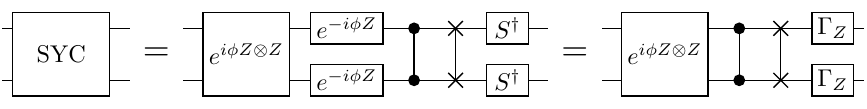}
    \caption{Circuit decomposition of the $\syc$ gate: $\Gamma_Z = S^\dagger e^{-i\phi Z} = e^{-i\phi Z} S^\dagger$ and $\phi = -\pi/24$, where two solid dots linked by a line represent the \textsc{cz} gate and two crosses linked by a line represent the \textsc{swap} gate.}
    \label{fig:supremacy}
\end{figure}

\begin{figure}[htbp!]
    \includegraphics[width=0.9\textwidth]{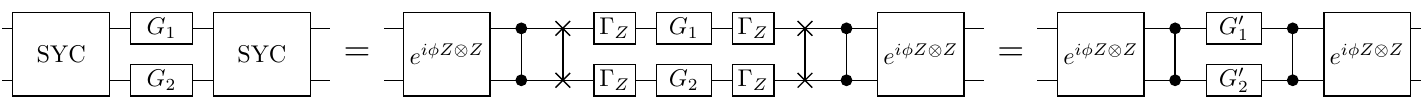}
    \caption{Single-qubit gates sandwiched by two $\syc$ gates:  The $\Gamma_Z$ gates map single-qubit operations to single-qubit operations}
    \label{fig:two_supremacies}
\end{figure}

The $\syc$ gate is an fSim($\pi/2$,$\pi/6$) which can be broken down into a $\textsc{cphase}(\pi/6)$, \textsc{cz}, \textsc{swap}, and two S gates according to \fig{supremacy}.
We analyze the KAK coefficients for a composite gate of two $\syc$ gates sandwiching arbitrary single qubit rotations, depicted in \fig{two_supremacies}, to determine the space of gates accessible with two $\syc$ gates.

Any two qubit gate is locally equivalent to standard KAK form~\cite{PhysRevA.67.042313}.  The coefficients in the KAK form is equivalent to the operator Schmidt coefficients of the 2-qubit unitary.  To find the Schmidt coefficients, we introduce the matrix representation of 2-qubit gates in terms of Pauli operators, i.e., the $jk$-th matrix element equals to the corresponding coefficient of the Pauli operator $P_j\otimes P_k$, where $P_{0, 1, 2, 3} = I, X, Y, Z$,
\begin{align}
    O_M = \sum_{j,k=0}^3 M_{jk} P_j\otimes P_k\,.
\end{align}
The Schmidt coefficients of $O_M$ equal to the singular values of $M$.  Any single-qubit gate $G_{1,2}'$ can be decomposed into the $Z$-$X$-$Z$ rotations; the $Z$ rotations commute with the \textsc{cz} and the \textsc{cphase}, and they do not affect the Schmidt coefficients of the two-qubit operation defined in \fig{two_supremacies}.  We neglect the $Z$ rotations and simplify $G_{1,2}'$ to single-qubit $X$ rotations
\begin{gather}\label{eq:single-qubit_final}
G_1' = \cos\theta_1 I + i\sin\theta_1 X\,,\qquad
G_2' = \cos\theta_2 I + i\sin\theta_2 X\,.
\end{gather}
The Pauli matrix representation of $G_1'\otimes G_2'$ in Eq.~\eqref{eq:single-qubit_final} is
\begin{align}
    A =
    \begin{pmatrix}
        c_1 c_2 & i c_1 s_2 & 0 & 0\\
        i s_1 c_2 &  -s_1 s_2 & 0 & 0\\
        0             & 0             & 0 & 0\\
        0 & 0 & 0 & 0
    \end{pmatrix}\,,
\end{align}
where $c_{1,2} = \cos\theta_{1,2}$ and $s_{1,2} = \sin\theta_{1,2}$.  The rank of the matrix $A$ is one, representing a product unitary.  After being conjugated by the \textsc{cz} gates, i.e, $O \mapsto \textsc{cz} \, O \, \textsc{cz}$, the matrix $A$ becomes
\begin{align}
    A\mapsto B =
    \begin{pmatrix}
        c_1c_2 & 0 & 0 & 0\\
        0 & 0 & 0 & is_1c_2\\
        0  & 0 & -s_1s_2 & 0\\
        0 & i c_1 s_2 & 0 & 0
    \end{pmatrix}\,,
\end{align}
where we use the relations for $O\mapsto \textsc{cz}\, O \, \textsc{cz}$,
\begin{align}
    X_1 X_2 \mapsto Y_1 Y_2\,,\quad X_1  \mapsto X_1 Z_2\,,\quad  X_2 \mapsto Z_1 X_2\,. 
\end{align}
The \textsc{cphase} part in the $\syc$ gate is
\begin{align}
    e^{i\phi Z\otimes Z} = \cos\phi\, I\otimes I + i\sin\phi\, Z\otimes Z \,,
\end{align}
where $\phi = -\pi/24$.  An arbitrary operator $O$ left and right multiplied by $\cphase$ part is expressed as
\begin{align}
    e^{i\phi Z\otimes Z} O e^{i\phi Z\otimes Z}  = (\cos\phi)^2 O + \frac{i}{2}\sin(2\phi) \left( Z^{\otimes 2} O + O Z^{\otimes 2}\right) - (\sin\phi)^2 Z^{\otimes 2} O Z^{\otimes 2}\,.
\end{align}
Applying the operation $O \mapsto \frac{1}{2} (Z^{\otimes 2} O + O Z^{\otimes 2})$ to the operator $B$, we have
\begin{align}
    B\mapsto C =
    \begin{pmatrix}
        0 & 0 & 0 & 0\\
        0 & s_1 s_2 & 0 & 0\\
        0             & 0             & 0 & 0\\
        0 & 0 & 0 & c_1  c_2
    \end{pmatrix}\,.
\end{align}
Applying the operation $O \mapsto Z^{\otimes 2} O Z^{\otimes 2}$ to the operator $B$, we have
\begin{align}
    B\mapsto D =
    \begin{pmatrix}
        c_1 c_2 & 0 & 0 & 0\\
        0 & 0 & 0 & -i s_1c_2\\
        0  & 0   & -s_1s_2 & 0\\
        0 & -ic_1s_2 & 0 & 0
    \end{pmatrix}\,.
\end{align}
The resulting two-qubit gate at the output of the circuit in \fig{two_supremacies} takes the form
\begin{align}
    M = (\cos\phi)^2 B + i\sin(2\phi)\, C - (\sin\phi)^2 D\,.
\end{align}
Two singular values of $M$ are $\cos(2\phi) c_1 c_2$ and $\cos(2\phi) s_1 s_2$ corresponding to the diagonal matrix elements $M_{0,0}$ and $M_{2,2}$, and the magnitudes of these two singular values are bounded by the angle $\phi$. Consider the two-dimensional subspace of the matrix $B$, $C$, and $D$ with the two known singular values removed
\begin{align}
    B\mapsto B' =
    \begin{pmatrix}
        0 & i s_1 c_2\\
        i c_1 s_2 & 0
    \end{pmatrix}\,,\quad
    C\mapsto C' =
    \begin{pmatrix}
        s_1 s_2  & 0\\
        0 &  c_1 c_2
    \end{pmatrix}\,,\quad
    D\mapsto D' =
    \begin{pmatrix}
        0 & -i s_1 c_2\\
        -i c_1 s_2 & 0
    \end{pmatrix}
\end{align}
The Pauli representation matrix in the reduced space is
\begin{align}
    M' &= (\cos\phi)^2 B' + i\sin(2\phi)\, C' - (\sin\phi)^2 D'\\[3pt]
    &=
    i
    \begin{pmatrix}
        \sin(2\phi) s_1 s_2 &  s_1 c_2\\[3pt]
        c_1 s_2 & \sin(2\phi) c_1 c_2
    \end{pmatrix}
    =
    i c_1 c_2
    \begin{pmatrix}
        \sin(2\phi) t_1 t_2 &  t_1\\[3pt]
        t_2 & \sin(2\phi)
    \end{pmatrix}\,.
\end{align}
To calculate the singular values of a $2\times 2$ matrix
\begin{align}
    M_{\boldsymbol\alpha} = \alpha_0  I + \alpha_1 X +\alpha_2 Y +\alpha_3 Z \,,
\end{align}
we used the formula
\begin{align}
    \sigma_\pm &= \sqrt{\eta \pm \sqrt{\eta^2 - \xi^2}}\,,
\end{align}
where $\eta = |\alpha_0|^2 + |\alpha_1|^2 + |\alpha_2|^2 + |\alpha_3|^2$ and $\xi = |\alpha_0^2 - \alpha_1^2 - \alpha_2^2 - \alpha_3^2|$.  For matrix $M'$, we have,
\begin{align}
    \eta &= \frac{1}{2}\,\sum_{j,k}\, \lvert M'_{jk}\rvert^2\\
    &= \frac{1}{2}\,\left(\sin(2\phi)^2 s_1^2 s_2^2 + s_1^2 c_2^2 + c_1^2 s_2^2 + \sin(2\phi)^2 c_1^2 c_2^2\right)\\[3pt]
    & = \frac{1}{2} - \frac{1}{2} \cos(2\phi)^2 \left(s_1^2 s_2^2 + c_1^2 c_2^2\right)\,.
\end{align}
and
\begin{align}
    \xi &= \frac{1}{4}\,\Big\lvert\sin(2\phi)^2\left(s_1s_2 + c_1c_2\right)^2 - ( s_1c_2 + c_1s_2)^2 + ( s_1 c_2 - c_1 s_2)^2 - \sin(2\phi)^2\left(s_1s_2 - c_1 c_2\right)^2\Big\rvert\\[2pt]
    &= \cos(2\phi)^2\, \big\lvert s_1 s_2  c_1c_2   \big\rvert\,.
\end{align}
We have solved all the four singular values of the 2-qubit unitary at the output of \fig{supremacy},
\begin{align}
    \lambda_0 = \vert \cos(2\phi) c_1 c_2\vert, \;\;\lambda_1 = \vert \cos(2\phi) s_1 s_2\vert, \;\;\lambda_2 = \sqrt{\eta + \sqrt{\eta^2 - \xi^2}},\;\;\lambda_3 = \sqrt{\eta - \sqrt{\eta^2 - \xi^2}}.
\end{align}
For the case $s_1 = 0$ and $c_1 = 1$, we have $\lambda_1 = \lambda_3 = 0$ and the other two singular values
\begin{align}
    \lambda_0 = \vert \cos(2\phi)c_2\vert \in [0,\;\cos(2\phi)] \,,\quad \lambda_2 = \sqrt{2\eta} = \sqrt{1 - \cos(2\phi)^2 c_2^2}\;.
\end{align}
Since $\cos(2\phi) \simeq 0.966 > 1/\sqrt 2$, we can implement any \textsc{cphase} gate using only two $\syc$ gates.  This is achieved by matching the Schmidt coefficients of $e^{-i \theta ZZ/2}$ to $\lambda_{0}$ and $\lambda_{2}$.  If $\vert \cos(\theta)\vert > \cos(2 \phi)$ then we can reset $c_{1,2}$ and $s_{1,2}$ appropriately to select out the other pair of singular values.

\begin{figure}[H]
    \centering
    \includegraphics[width=0.6\textwidth]{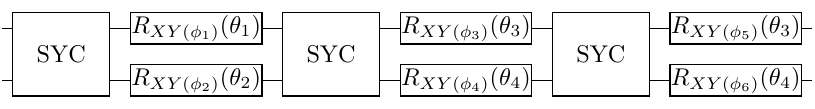}
    \caption{Circuit used to match the KAK coefficients of the \textsc{swap} gate. The $R_{XY}(\phi)(\theta)$ is a rotation of $\theta$ around an axis in the $XY$-plane defined by $\phi$. This is implemented in Cirq as a PhasedXPow gate.}
    \label{fig:swap_kak_matching}
\end{figure}

\textbf{Compilation of $\swap$}.
A $\swap$ gate requires three applications of $\syc$ and is used for the 3-regular MaxCut problem circuits.  The $\swap$ gate was numerically compiled by optimizing the angles of the circuit in \fig{swap_kak_matching} to match the KAK interaction coefficients for the $\swap$ gate.
After the the angles in the circuit depicted in \fig{swap_kak_matching} are determined to match the KAK coefficient of the swap gate we add single qubit rotations to make the circuit fully equivalent to $\swap$.

\textbf{Compilation of $\zzswap$}.
This composite interaction can be effected with three applications of $\syc$ and is used for SK-model circuits.  The $\syc$ gate KAK coefficients are $(\pi/4, \pi/4, \pi/24)$ which is locally equivalent to a $\cphase(\pi/4 - \pi/24)$ followed by a $\swap$.
Therefore, to implement a $ZZ(\gamma)$ followed by a swap we need to apply a single $\syc$ gate followed by the composite $\cphase(\gamma - \pi/24 + \pi/4)$.  The total composite gate now involves 3 $\syc$ gates, a single $Rx$ gate and two $Rz$ gates.

\textbf{Scheduling of Hardware Grid gates}. An efficient planar graph edge-coloring can be used to schedule as many simultaneous $ZZ$ interactions as possible. We activate links on the graph in the following order:
1) horizontal edges starting from even nodes;
2) horizontal edges starting from odd nodes;
3) vertical edges starting from even nodes;
4) vertical edges starting from odd nodes.
Viewed as cardinal directions and choosing an even node as the central point this corresponds to a west, east, south, north (W, E, S, N) activation sequence. \\

\begin{figure}[!htbp]
    \centering
    \includegraphics[width=4in]{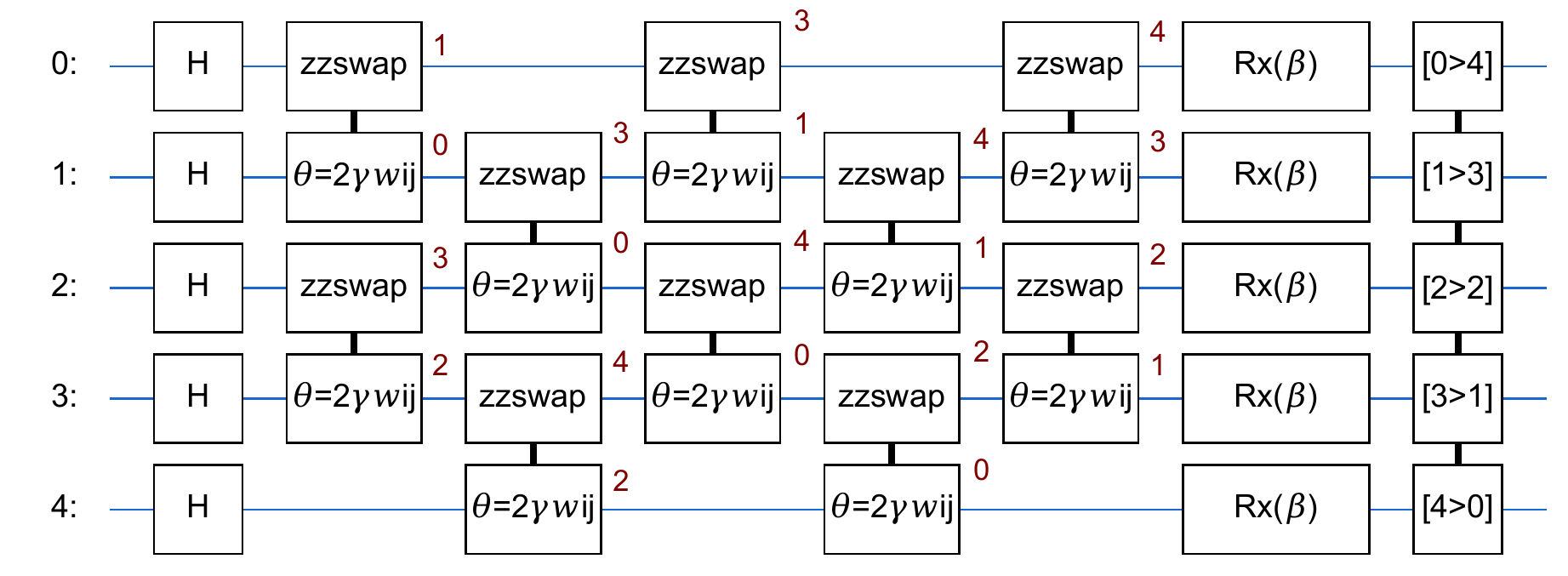}
    \caption{$p=1$ swap network for a 5-qubit SK-model.
    Physical qubits are indicated by horizontal lines and logical node indices are indicated by red numbers.
    The network effects all-to-all logical interactions with nearest-neighbor interactions in depth $n$.}
    \label{fig:swap-network}
\end{figure}

\textbf{Fully Connected Swap Network}.
All-to-all interactions can be implemented
optimally with a swap network in which pairs of linear-nearest-neighbor qubits are repeatedly interacted and swapped.
Crucially, the required interactions $\swap$ and $e^{-i \gamma ZZ}$ between all pairs all mutually commute so we are free to re-order all two-qubit interactions to minimize compiled circuit depth.
After $n$ applications of layers of $\zzswap$ interactions (alternating between even and odd qubits), every qubit has been involved
in a $ZZ$ interaction with every other qubit and logical qubit indices have been reversed.
This can be viewed as a (parallel) bubble sort algorithm initialized with a reverse-sorted list of logical qubit indices. An example at $n=5$ is shown in \fig{swap-network}. If $p$ is even, two applications of the swap network return qubit indices to their original mapping. Otherwise, post-processing can reverse the measured bitstrings.

\begin{figure}[!htbp]
    \centering
    \includegraphics[width=2in]{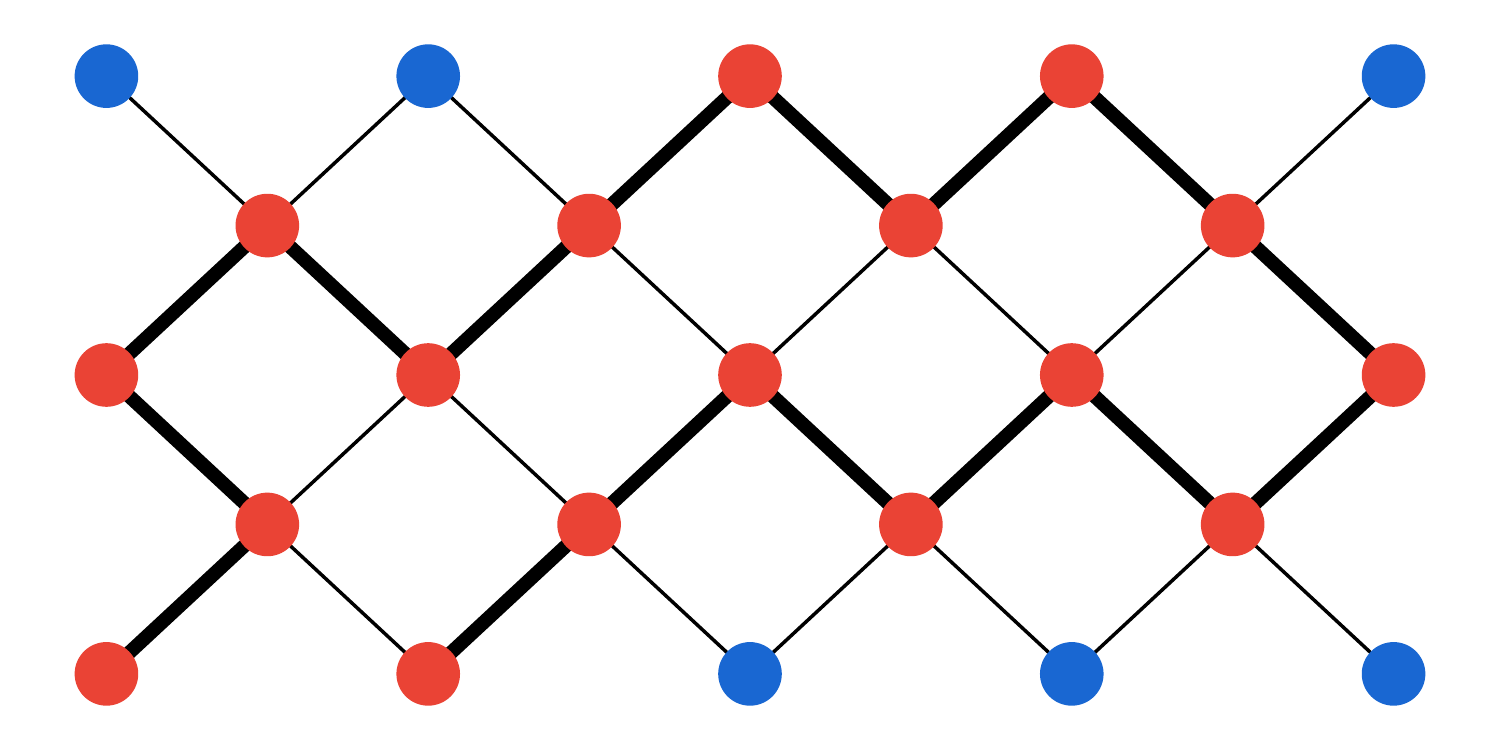}
    \caption{The largest line one can embed on the 23-qubit device is of length 17.}
    \label{fig:snake-line}
\end{figure}

The swap network requires linear connectivity. On the 23-qubit subgraph of the Sycamore device used for this experiment, this limits us to a maximum size of $n=17$ for the SK model, shown in \fig{snake-line}.

\section{Prior Work}
\label{app:compare}

\begin{table*}[htbp]
    \begin{tabular}{l|l|l|c|c|c|c}
        Reference
        & Date
        & Problem topology
        & $\Delta(G)$
        & $n$
        & $p$
        & Optimization \\
        \hline\hline
        \citet{rigetti-qaoa-clustering}
        & 2017-12 \quad
        & Hardware
        & 3
        & 19
        & 1
        & Yes \\
%
        \citet{2018-bristol-photonics}
        & 2018-08
        & Hardware
        & 1
        & 2
        & 1
        & No \\
%
        \citet{2019-monroe-qaoa}
        & 2019-06
        & Hardware\textsuperscript{1} (system 1)
        & $n$
        & 12, 20
        & 1
        & Yes \\
%
        & 
        & Hardware\textsuperscript{1} (system 2)
        & $n$
        & 20--40
        & 1--2\textsuperscript{(2)}
        & No \\
%
        \citet{2019-julich-qaoa-on-ibm}
        & 2019-07
        & Hardware
        & 3
        & 8
        & 1
        & No \\
%
        \citet{rigetti-xy-gate}
        & 2019-12
        & Ring
        & 2
        & 4
        & 1
        & No \\
%
        & %
        & Fully-connected
        & $n$
        & 
        & 
        & No \\
%
        \citet{chalmers-2q-qaoa}
        & 2019-12
        & Hardware
        & 1
        & 2
        & 1, 2
        & Yes \\
%
        This work
        & %
        & Hardware
        & 4
        & 2--23
        & 1--5
        & Yes \\
        &
        & 3-regular
        & 3
        & 4--22
        & 1--3
        & Yes \\
%
        & %
        & Fully-connected
        & $n$
        & 3--17
        & 1--3
        & Yes \\
        \hline
    \end{tabular}
    \caption{
    An overview of experimental demonstrations of QAOA. Although each work generally frames the algorithm in terms of a combinatorial optimization problem (2SAT, Exact Cover, etc.), we classify problems based on their topology, maximum degree of the problem graph $\Delta(G)$, the number of qubits $n$ and the depth of the algorithm $p$. These attributes give a rough view of the difficulty of a particular instance. We indicate whether variational optimization of parameters was demonstrated. \textsuperscript{1}In superconducting processors, ``Hardware" topologies are 2-local planar lattices. In ion trap processors, hardware-native topologies are long range couplings of the form $J_{ij} \approx J_0/|i - j|^\alpha$.
    \textsuperscript{2}$p=2$ only for $n=20$.
    }
\end{table*}

Prior work has included experimental demonstration of the QAOA. The referenced works often include additional results, but we focus specifically on the sections dealing with experimental implementation of the algorithm.

\citet{rigetti-qaoa-clustering} demonstrated a Bayesian optimization of p = 1 parameters on a 19-bit hardware-native Ising graph using a Rigetti superconducting qubit processor. The authors compared the cumulative probability of finding the lowest energy bitstring over the course of the optimization to binomial coin flips and showed performance from the device exceeding random guessing. The problem topology involved a roughly hexagonal tessellation. The problems were related to a restricted form of two-class clustering.

\citet{2018-bristol-photonics} demonstrated a $n=2$,  $p=1$ QAOA landscape on their photonic quantum processor. They presented three instances of the two-bit problem, which was framed as Max2Xor. The color scale for the landscapes was re-scaled for experimental values. They demonstrated high probability of obtaining the correct bitstrings.

\citet{2019-monroe-qaoa} demonstrated application of the QAOA with two ion trap quantum processors, called ``system 1" and ``system 2". The problems were of the form $J_{ij} \approx J_0 / |i - j|^\alpha$ with $\alpha$ close to unity. This corresponds to an antiferromagnetic 1D chain. This problem is fully connected, but is spiritually similar to the hardware native planar graphs studied in superconducting architectures in the sense that the cost function cannot be programmed and is easily solvable at any system size. A landscape is shown for $n = 20$ from system 1. Optimization traces are shown for $n = 12$ and $n = 20$ on system 1. On system 2, performance was demonstrated at optimal parameters for $n = \{20, 25, 30, 35, 40\}$. Additionally, a partial $p = 2$ grid search was performed on system 2. Nine discrete choices for $(\gamma_1, \beta_1, \beta_2)$ were selected and then a scan over $\gamma_2$ was reported for each choice. Finally, on system 2, performance was compared between $p = 1$ and $p = 2$ at $n = 20$, giving a ratio of $(93.8 \pm 0.4)\%$ versus $(93.9 \pm 0.3)\%$, respectively.

\citet{2019-julich-qaoa-on-ibm} demonstrated an application of the QAOA via IBM’s Quantum Experience cloud service on the 16Q Melbourne device. The 8-bit problem studied was framed as 2SAT and had a topology matching the device with maximum node degree of 3. A landscape with re-scaled color map was compared to the theoretical landscape.

\citet{rigetti-xy-gate} implemented QAOA on two types of problems; each with two compilation strategies. The 4-bit problems had a ring topology and a fully-connected topology. While a 4-qubit ring would fit on the Rigetti superconducting device, they implemented both problems using only linear connectivity with the introduction of $\swap$s. In one compilation strategy, they used $\textsc{cz}$ as the gate-synthesis target. In the other, they used both $\textsc{cz}$ and $\iswap$. The color bars were re-scaled for the experimental data.

\citet{chalmers-2q-qaoa} ran 2-bit QAOA instances on their superconducting architecture at $p = 1$ and $p = 2$. They show four $p = 1$ landscapes and demonstrate optimization for $n = 2, p = 2$. They observed that increasing circuit depth to $p = 2$ increases the probability of observing the correct bitstring.

\section{Readout correction}\label{app:readout}
The experimentally measured expectation values plotted in \fig{landscapes} were adjusted with a procedure
used to compensate for qubit readout error. We model readout error as a classical bit-flip error channel that changes the
measurement result of qubit $i$ from 0 to 1 with probability $p_{0, i}$ and from 1 to 0 with probability $p_{1, i}$.
Under the effect of this error channel, a measurement of a single qubit in the computational basis is described by the following
positive operator-valued measure (POVM) elements (we drop the subscript $i$ here for clarity):
\begin{align}
    \tilde{\Pi}_0 &= (1 - p_0) \Pi_0 + p_1 \Pi_1 \\
    \tilde{\Pi}_1 &= p_0 \Pi_0 + (1 - p_1) \Pi_1,
\end{align}
where $\Pi_0 = \ket{0}\bra{0}$, $\Pi_1 = \ket{1}\bra{1}$. The uncorrected $Z$ observable
can be written as
\begin{align}
    \tilde{Z} = \tilde{\Pi}_0 - \tilde{\Pi}_1 = (p_1 - p_0)I + (1 - p_1 - p_0)Z.
\end{align}
Solving for $Z$, we have
\begin{align}
    Z = \frac{\tilde{Z} - (p_1 - p_0)I}{1 - p_1 - p_0}.
\end{align}
For our problems we are interested in the two-qubit observable $Z_i Z_j$, so the corrected observable is
\begin{align}
    Z_i Z_j = \frac{\tilde{Z_i} - (p_{1, i} - p_{0, i})I}{1 - p_{1, i} - p_{0, i}} \cdot \frac{\tilde{Z_j} - (p_{1, j} - p_{0, j})I}{1 - p_{1, j} - p_{0, j}}.
\end{align}
This expression tells us how to adjust the measured observable to compensate for the readout error. In the above analysis, we can replace $p_0$ and $p_1$ by their average $(p_0 + p_1) / 2$ if we perform measurements in the following way: for half of the measurements, apply a layer of $X$ gates immediately before measuring, and then flip the measurement results. In this case, the corrected observable is
\begin{align}
    Z_i Z_j = \tilde{Z_i}\tilde{Z_j} \cdot \frac{1}{1 - p_{1, i} - p_{0, i}} \cdot \frac{1}{1 - p_{1, j} - p_{0, j}}.
\end{align}
We estimated the value of $p_{0, i}$ on the device by preparing and measuring the qubit in the $\ket{0}$ state 1,000,000 times and counting how often
a 1 was measured; $p_{1, i}$ was estimated in the same way but by preparing the $\ket{1}$ state instead of the zero state. This estimation was performed periodically during the data collection for \fig{landscapes} to account for drift following automated calibration.

We measure each qubit via the state-dependent dispersive shift they induce on their corresponding harmonic readout resonator as described in \citet{google_supremacy_2019} supplementary information section III. We interrogate the readout resonator frequency with an appropriately calibrated microwave pulse (e.g. a frequency, power, and duration). When demodulated, the readout signal produces a `cloud’ of In-phase and Quadrature (IQ) Voltage points which are used to train an out state descrimator. Often, we find that optimal single-qubit calibrations extend to the case of simultaneous readout, but this is not always the case. For example, due to the Stark shift induced by photons in readout resonators, new frequency collisions may be introduced that are not present in the isolated readout case. Similarly, the combined power of a multiplexed readout pulses may exceed the saturation power of our parametric amplifier.

\begin{figure*}[!htbp]
    \centering
    \includegraphics[width=0.7\linewidth]{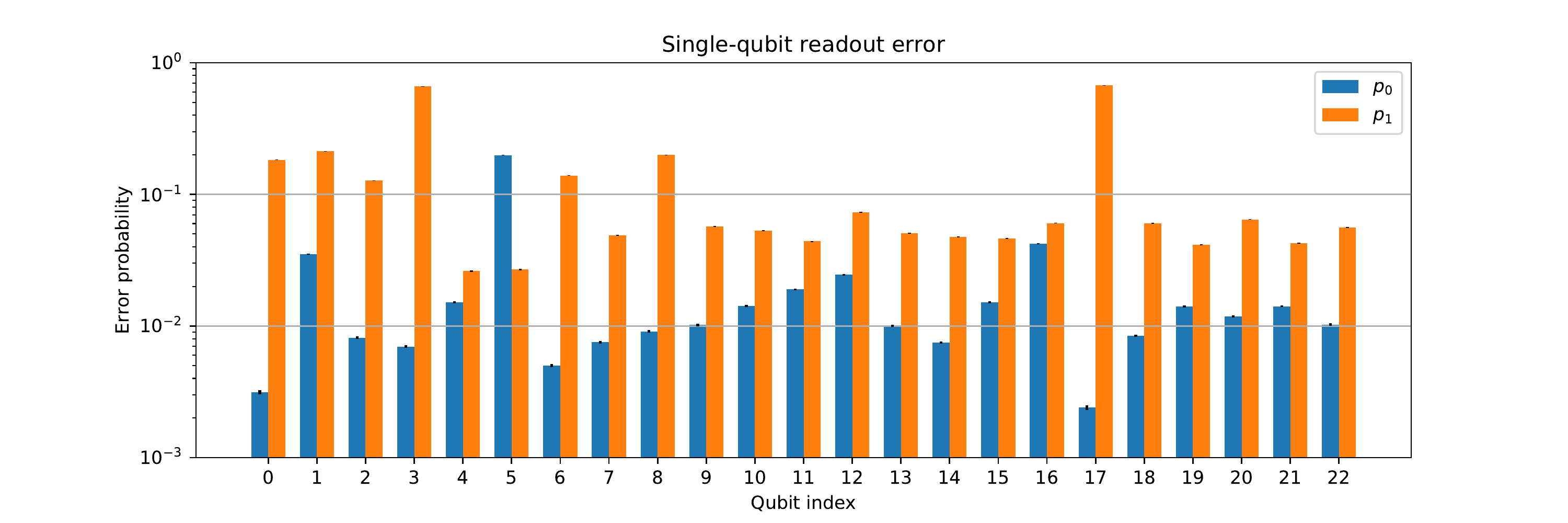}
    \includegraphics[width=0.7\linewidth]{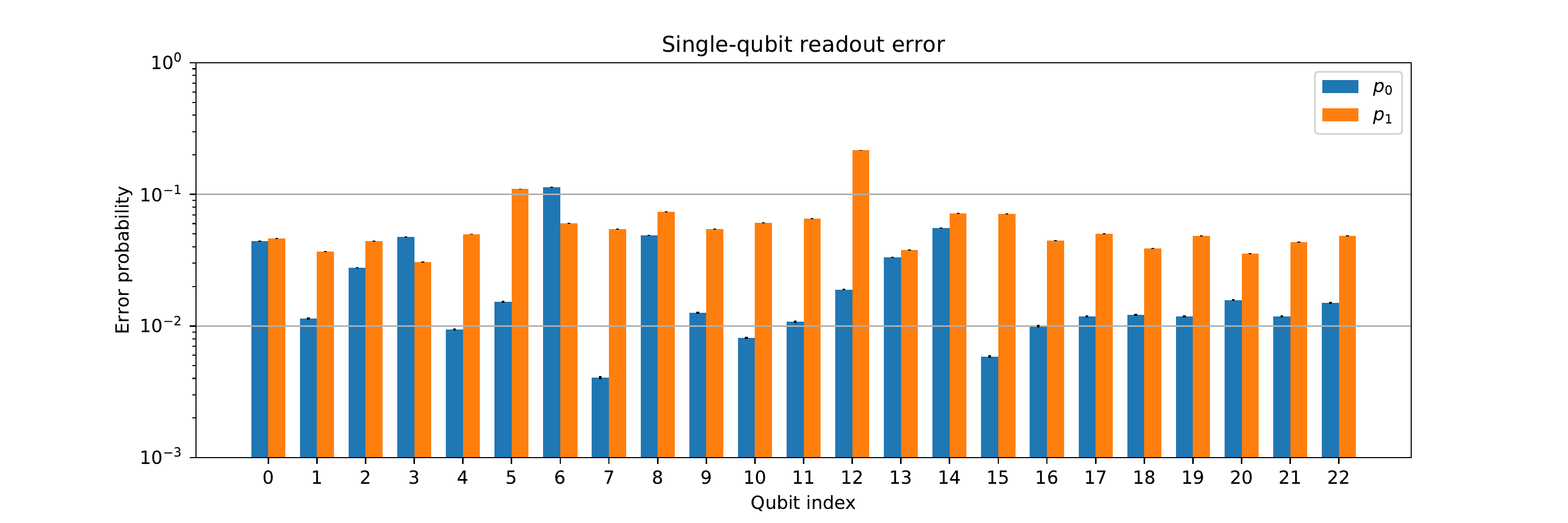}
    \caption{
    (Top) Marginalized error probabilities $p_{0, i}$ and $p_{1, i}$ for simultaneous readout of all qubits from a representative calibration used
    to correct \fig{landscapes} for readout error.
    (Bottom) Values for typical marginalized simultaneous readout error probabilities after the implementation of an improved automated calibration routine.
    Error bars (barely visible) represent a 95\% confidence interval.}
    \label{fig:readout-error}
\end{figure*}

At the time of the primary data collection for this experiment, all automated calibration routines were performed with each qubit in isolation. Subsequently, a calibration which optimizes qubit detunings during readout was implemented to mitigate these correlated readout errors caused by frequency collisions. \fig{readout-error} shows $\ket{0}$ and $\ket{1}$ state errors for \emph{simultaneous} readout of all 23 qubits (which are used to correct $\langle ZZ \rangle$ observables) both as they were during primary data taking for \fig{landscapes} (top) and after implementing the improved readout detuning calibration (bottom). During primary data collection, the median isolated readout error was 4.4\% as measured during the previous automated calibration.
The discrepancy between these figures and the calibration values shown in \fig{readout-error}, top can be attributed to drift since the automated system calibration in addition to the simultaneity effects described above.

Data presented in \fig{expect-val-hardware} and \fig{hardware-grid-p} was taken on a different date with median isolated readout error as 4.1\% as reported in the main text. Readout correction was not used for these two figures.

While automated calibrations will continue to improve, drift will likely remain an inevitability when controlling qubits with analog signals. As such, we expect the readout corrections employed here will continue to provide utility for end-users of cloud-accessible devices.
In general, there will always be a difference between a hands-on calibration conducted by an experimental physicist and automated calibration for a cloud-accessible device, and we look forward to future research ideas being “productionized” to make them accessible to a wide audience of algorithms researchers. Even in this instance there is still an imperfect abstraction: if one is interested in reading only a subset of all available qubits, higher performance can be obtained by doing a highly-tailored calibration; but we expect that the vast majority of cases will be served better by the new calibration routines.

\section{Optimizer Details}\label{app:optimizer}
In this section, we describe the classical optimization algorithm that we used to obtain the optimization results presented in \fig{landscapes}. The algorithm is a variant of gradient descent which we call ``Model Gradient Descent''. In each iteration of the algorithm, several points are randomly chosen from the vicinity of the current iterate. The objective function is evaluated at these points, and a quadratic model is fit to the graph of these points and previously evaluated points in the vicinity using least-squares regression. The gradient of this quadratic model is then used as a surrogate for the true gradient, and the algorithm descends in
the corresponding direction. Our implementation includes hyperparameters that determine the rate of descent, the radius of the vicinity from which points are sampled (the sample radius), the number of points to sample, and optionally, whether and how quickly
the rate of descent and the sample radius should decay as the algorithm proceeds. Pseudocode is given in \alg{mgd}.

\begin{algorithm}[H]
    \caption{Model Gradient Descent}\label{alg:mgd}
    \begin{algorithmic}[1]
        \Require Initial point $x_0$, learning rate $\gamma$, sample radius $\delta$, sample number $k$,
        rate decay exponent $\alpha$, stability constant $A$,
        sample radius decay exponent $\xi$, tolerance $\varepsilon$, maximum evaluations $n$
        \State Initialize a list $L$
        \State Let $x \leftarrow x_0$
        \State Let $m \leftarrow 0$
        \While{(\#function evaluations so far) + $k$ does not exceed $n$}
            \State Add the tuple $(x, f(x))$ to the list $L$
            \State Let $\delta' \leftarrow \delta / (m + 1)^\xi$
            \State Sample $k$ points uniformly at random from the $\delta'$-neighborhood of $x$;
            Call the resulting set $S$
            \For{each $x'$ in $S$}
                \State Add $(x', f(x'))$ to $L$
            \EndFor
            \State Initialize a list $L'$
            \For{each tuple $(x', y')$ in $L$}
                \If{$\lvert x' - x \rvert < \delta'$}
                    \State Add $(x', y')$ to $L'$
                \EndIf
            \EndFor
            \State Fit a quadratic model to the points in $L'$ using least squares
            linear regression with polynomial features
            \State Let $g$ be the gradient of the quadratic model evaluated at $x$
            \State Let $\gamma' = \gamma / (m + 1 + A)^\alpha$
            \If{$\gamma' \cdot \lvert g \rvert < \varepsilon$}
                \State \Return $x$
            \EndIf
            \State Let $x \leftarrow x - \gamma' \cdot g$
            \State Let $m \leftarrow m + 1$
        \EndWhile
        \State \Return $x$
    \end{algorithmic}
\end{algorithm}

\newpage
\section{Supporting Plots for Performance at Optimal Angles}\label{app:all-optimal-angle}
\subsection{Analysis of Noise}
There are two relevant mechanisms when considering the difference in performance between problems. One is the propagation of faults through the circuit and the other is fidelity decay due to circuit depth. A single fault on low-degree problems (Hardware Grid and 3-regular MaxCut, with degree four and three, respectively) can only propagate to terms $p$ edges away from the original location of the fault, irrespective of the total number of qubits. However, if compilation results in circuits extensive in the system size, the probability of a fault increases. For the SK-model, the degree of the problem is extensive in system size so both the propensity for fault propagation as well as the probability of faults grows with $n$. Additionally, compilation of the 3-regular problems onto the hardware topology introduces SWAPs, which can propagate faults through nodes which would otherwise not be adjacent in the problem graph.

\begin{figure*}[htbp!]
    \centering
    \includegraphics[width=0.9\textwidth]{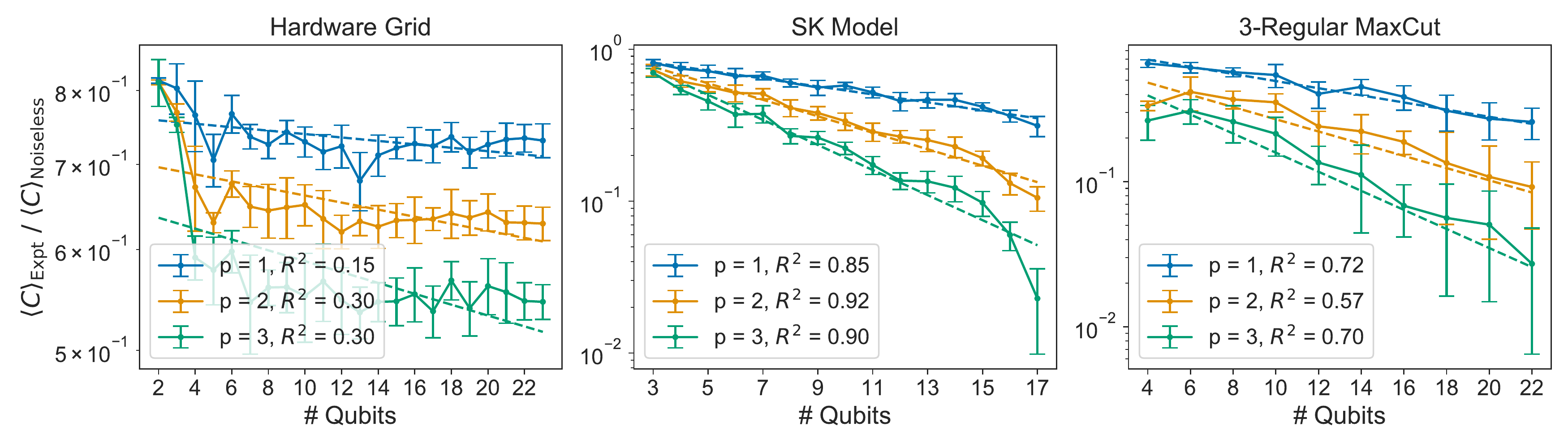}
    \caption{
    An exponential model compatible with a depolarizing error channel reasonably models the performance of compiled SK Model and 3-Regular MaxCut problems because their circuits are extensive in system size and faults are rapidly mixed. This error model is a poor fit for Hardware Grid problems due to the low degree of the problem graph and simple compilation.
    }
    \label{fig:fit-depol}
\end{figure*}

To probe these two effects, we fit a global depolarizing channel to the results for the three problems. A global depolarizing channel results in the mixed state $$
\rho = f_c |\psi\rangle\langle\psi| + \frac{1-f_c}{d} \mathbf{I}$$
where $|\psi\rangle$ is the noiseless QAOA state, $\mathbf{I}$ is the $n$-qubit identity matrix, and $f_c$ is the total circuit fidelity. $\mathrm{Tr}(\mathbf{I} C) = 0$ because of the $ZZ$ structure of the cost function, so the experimental objective function is simply a scaled version of the noiseless version, $\langle C \rangle_\mathrm{Expt} = f_c \langle C \rangle_\mathrm{Noiseless}$. We perform a linear regression on $f_c = f^n \times f_0 \leftrightarrow \log(f_c) = n \log(f) + \log(f_0)$ where $\log(f)$ and $\log(f_0)$ are fittable parameters physically corresponding to a per-qubit fidelity and a qubit-independent offset.
For the Hardware Grid, a depolarizing model is inappropriate, as the limited fault propagation and fixed circuit depth yield a largely $n$-independent noise signature. The exponential decay expected from a global depolarizing channel reasonably fits both the SK model and MaxCut results. We note that the fit is considerably stronger for the high-degree SK model where faults are rapidly mixed.

\newcommand{\spfpaoawidth}{2.8in}

\begin{figure*}[htbp!]
    \centering
    \includegraphics[width=\spfpaoawidth]{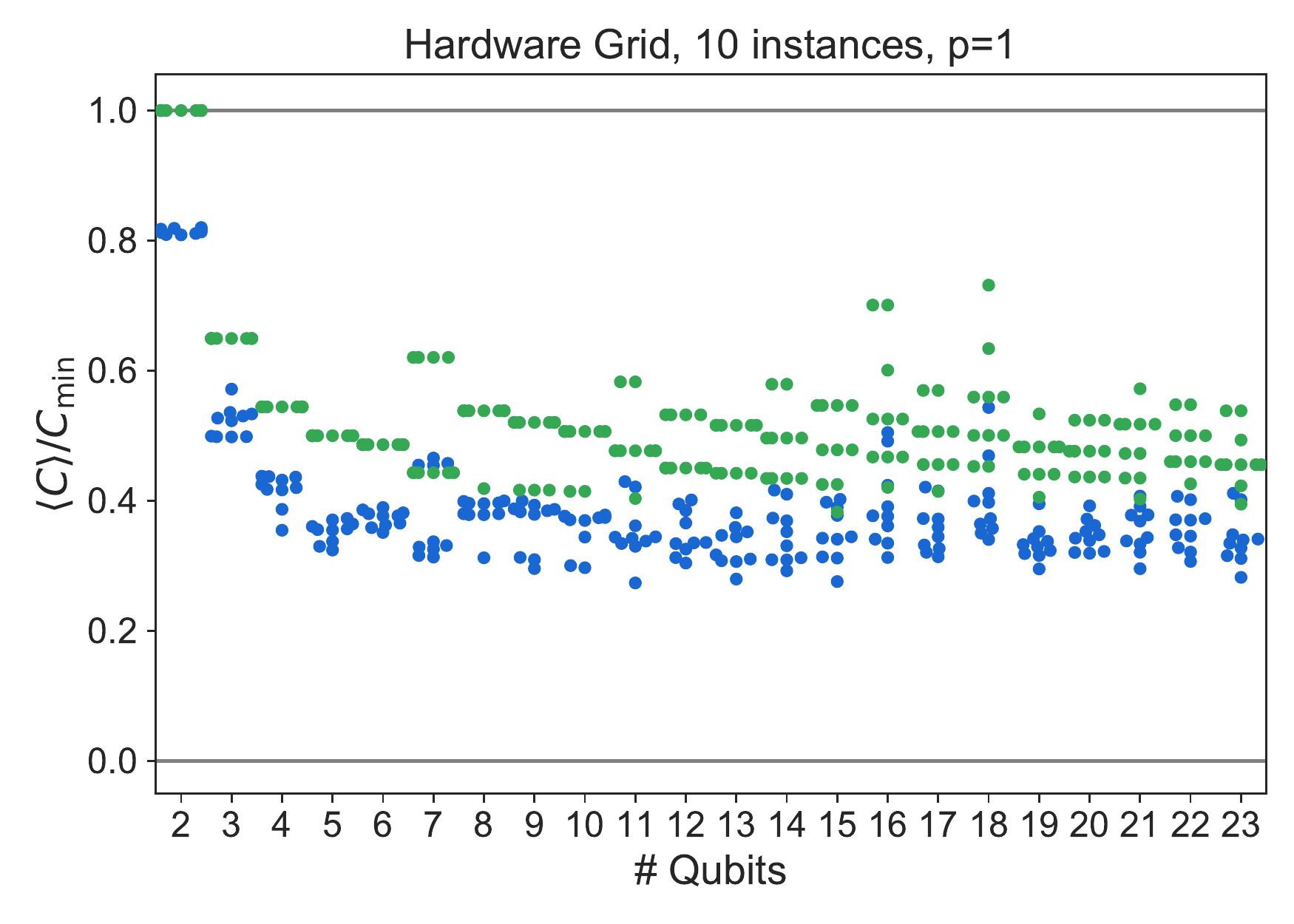}
    \includegraphics[width=\spfpaoawidth]{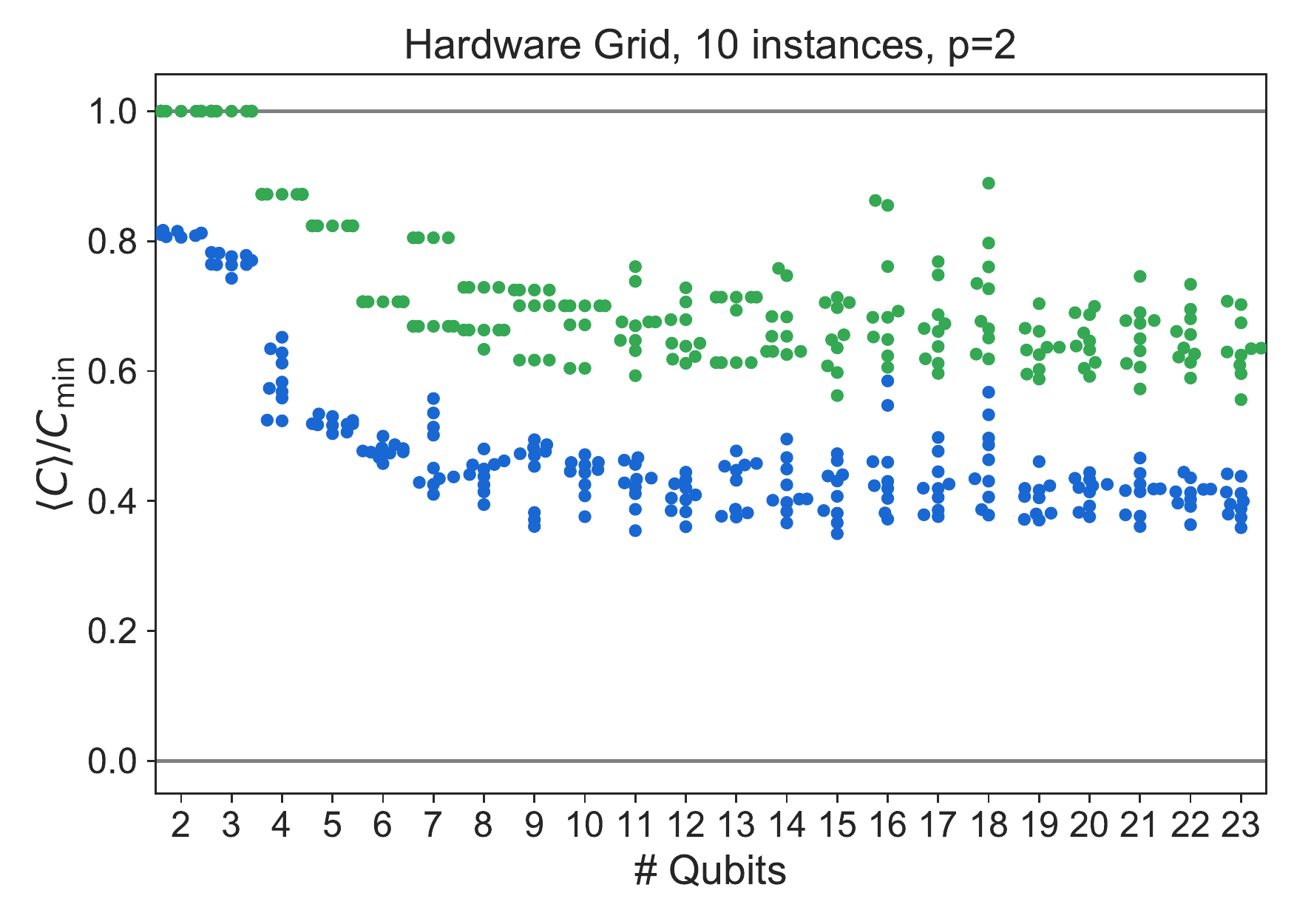}
    \includegraphics[width=\spfpaoawidth]{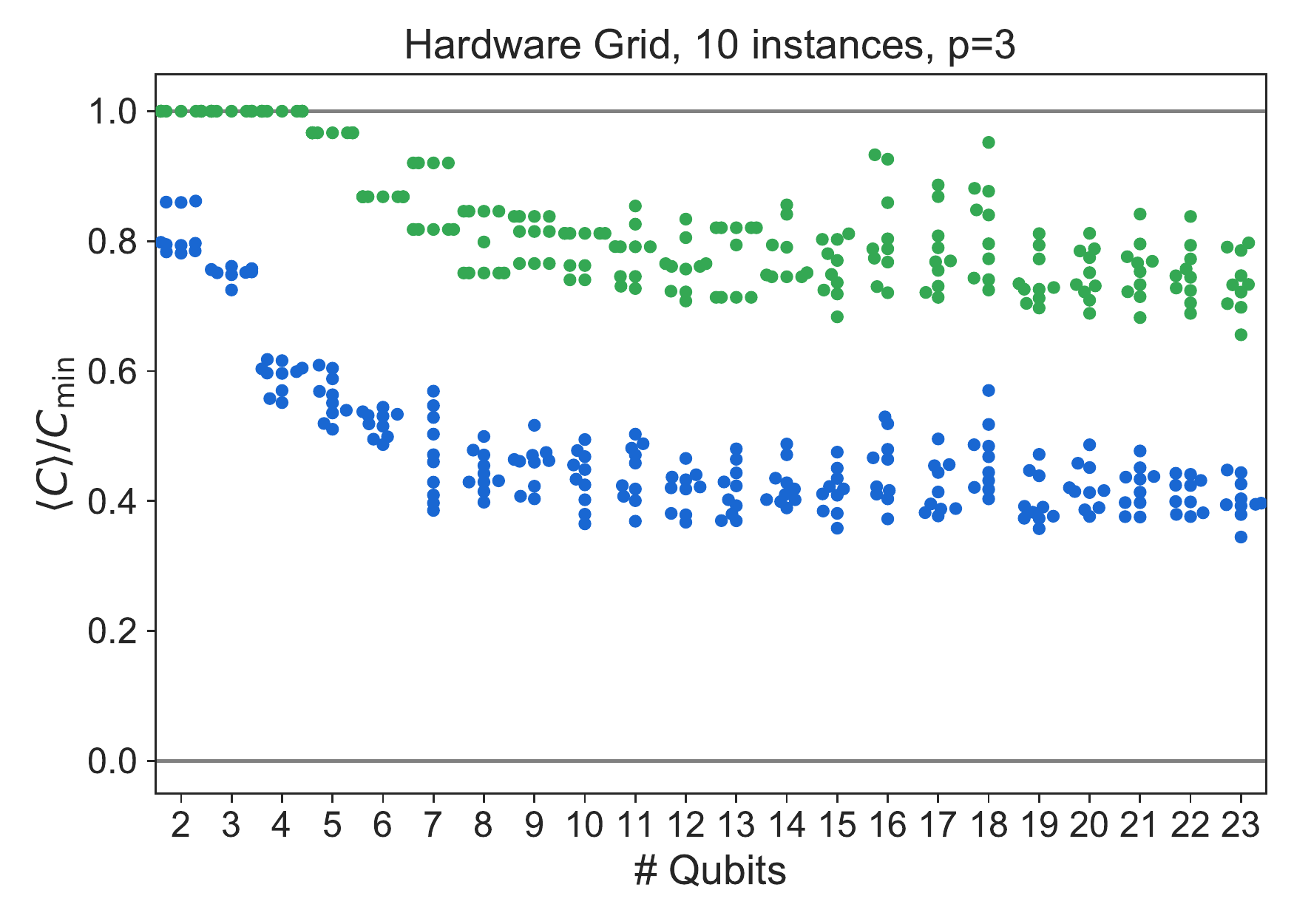}
    \includegraphics[width=\spfpaoawidth]{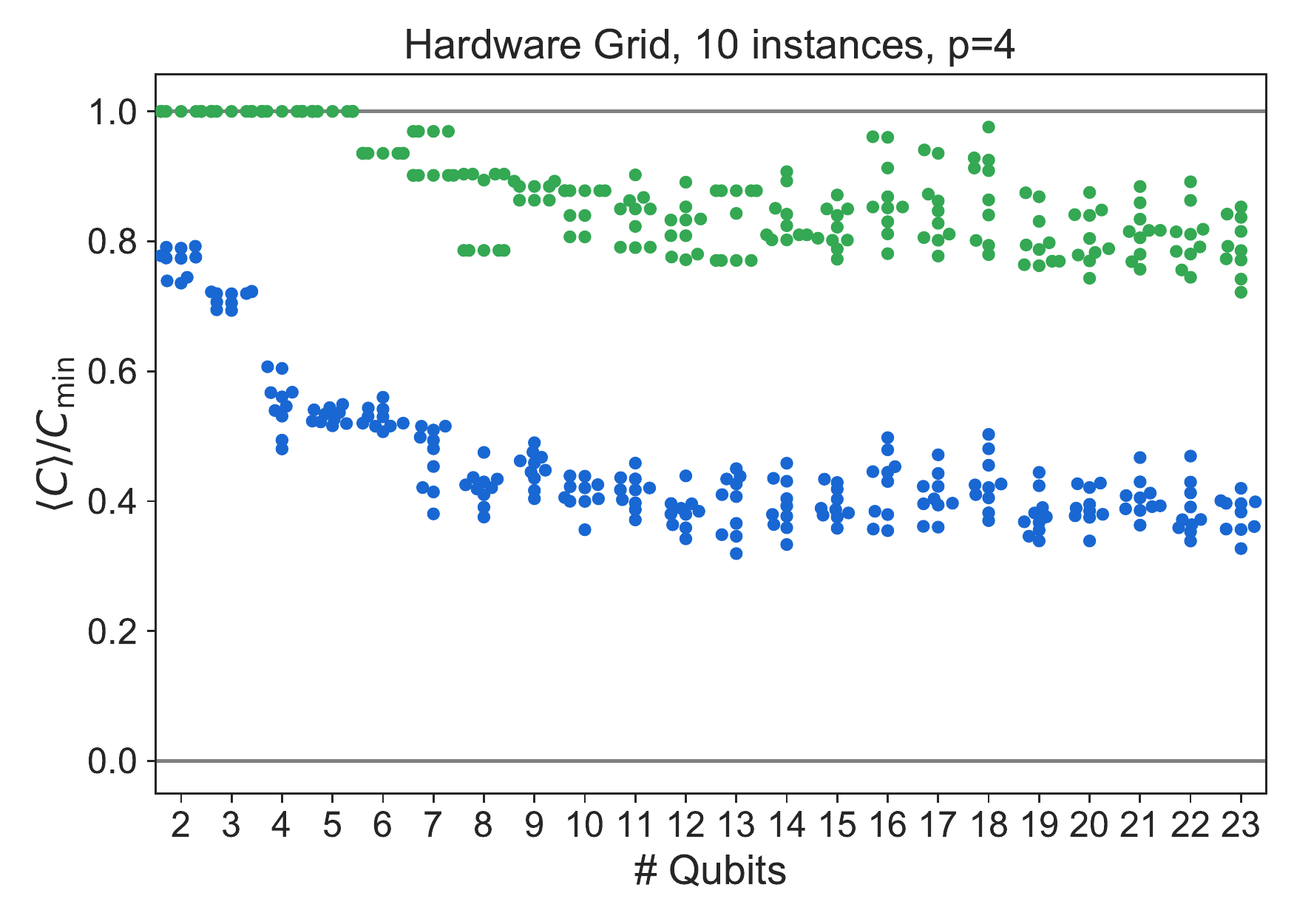}
    \includegraphics[width=\spfpaoawidth]{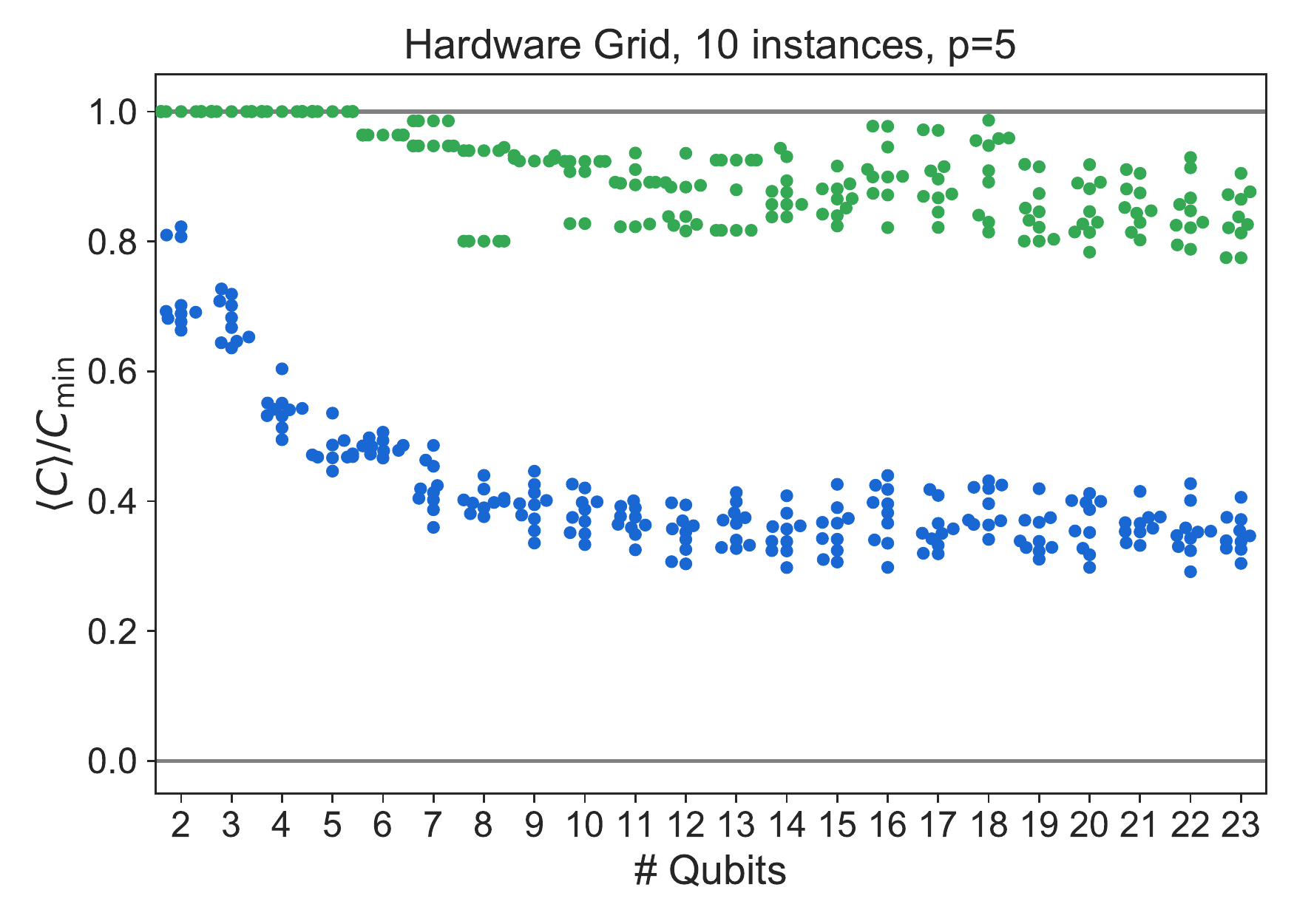}
    \caption{
    Performance of QAOA at $p \in [1, 5]$ and $n \in [2, 23]$ over random instantiations of couplings as described in the main text. Points have been perturbed along the $x$-axis to avoid overlap.
    \textbf{Green:} Noiseless
    \textbf{Blue:} Experimental
    }
    \label{fig:expectation-hardware-all}
\end{figure*}

\begin{figure*}[htbp!]
    \centering
    \includegraphics[width=\spfpaoawidth]{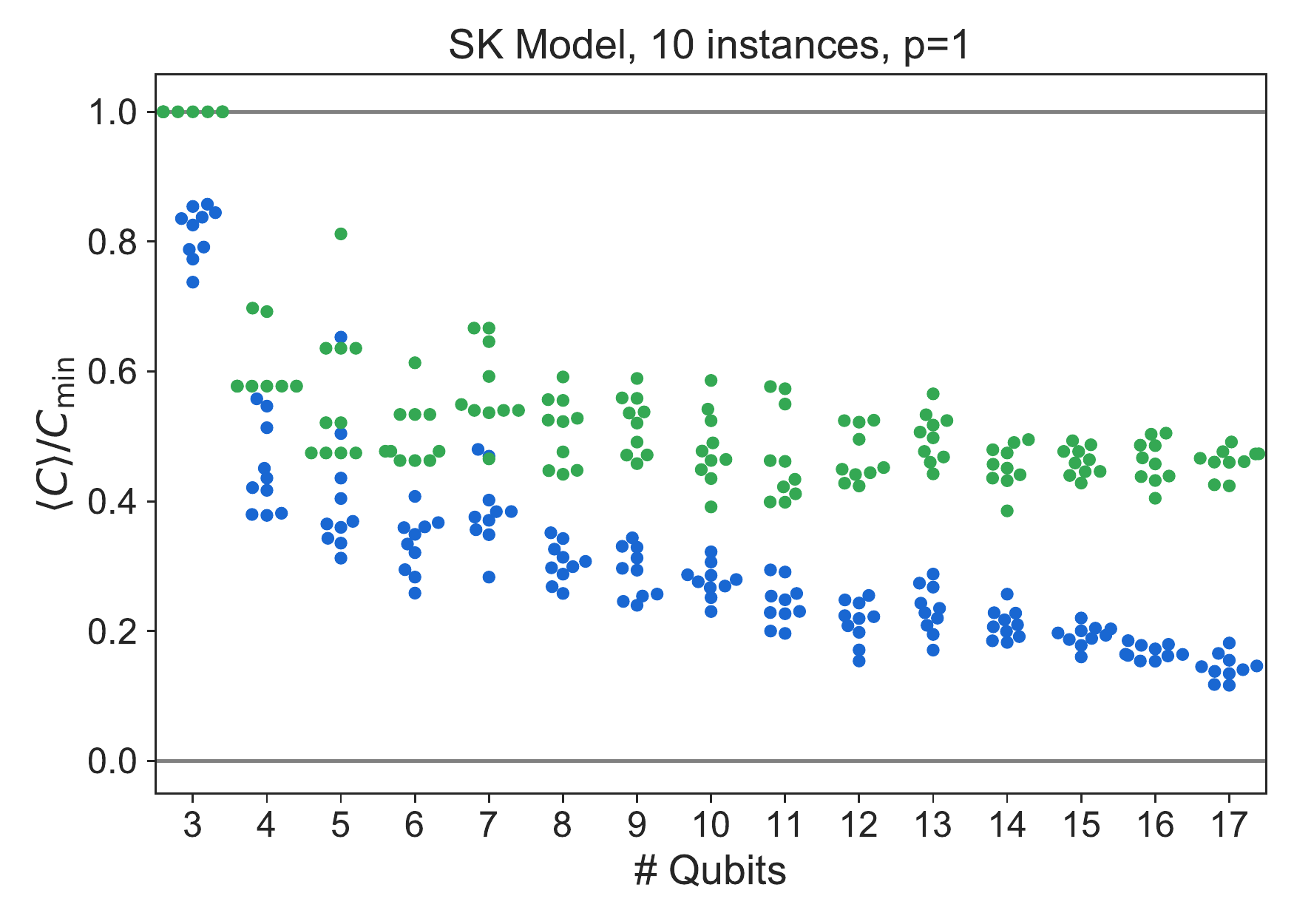}
    \includegraphics[width=\spfpaoawidth]{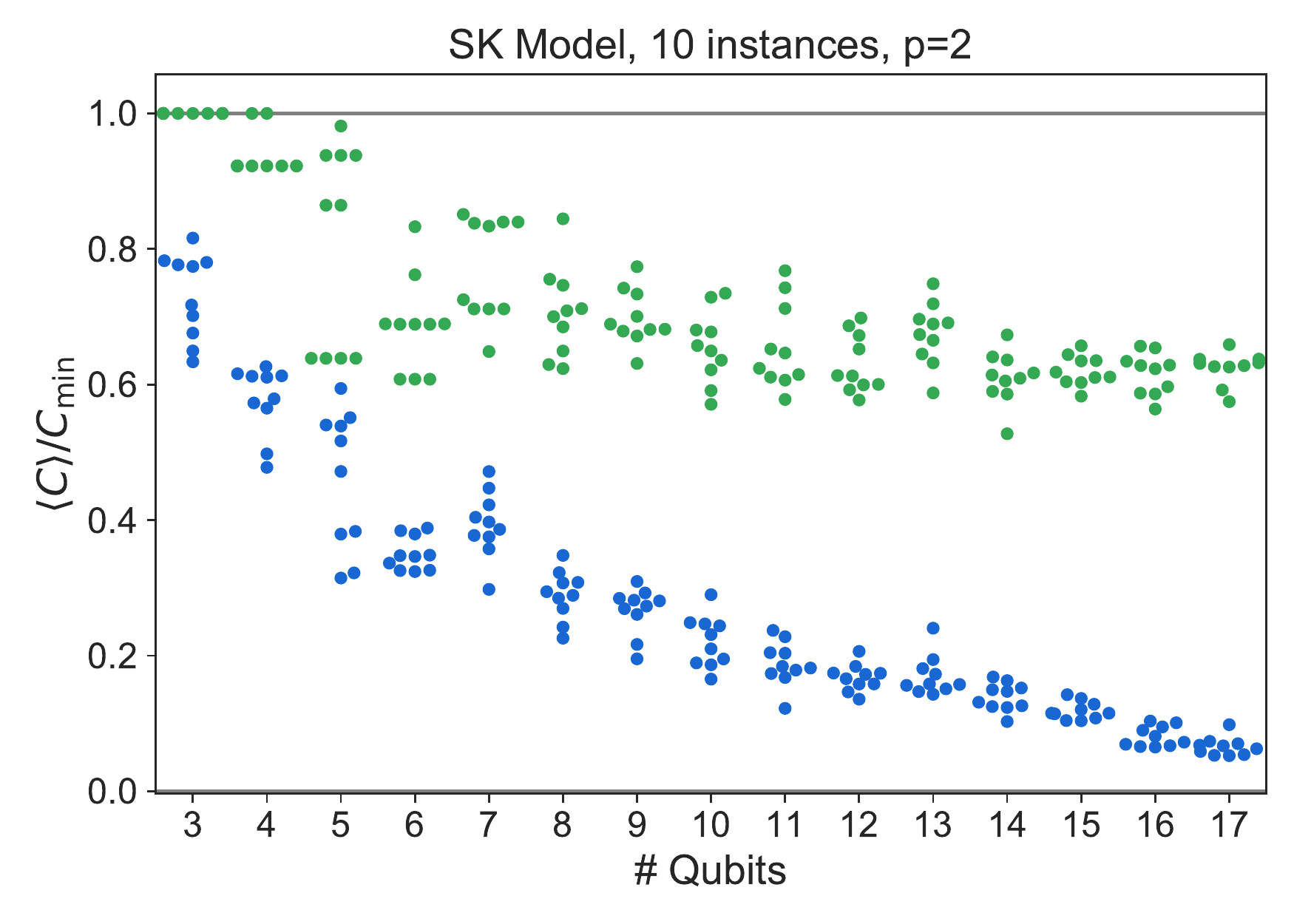}
    \includegraphics[width=\spfpaoawidth]{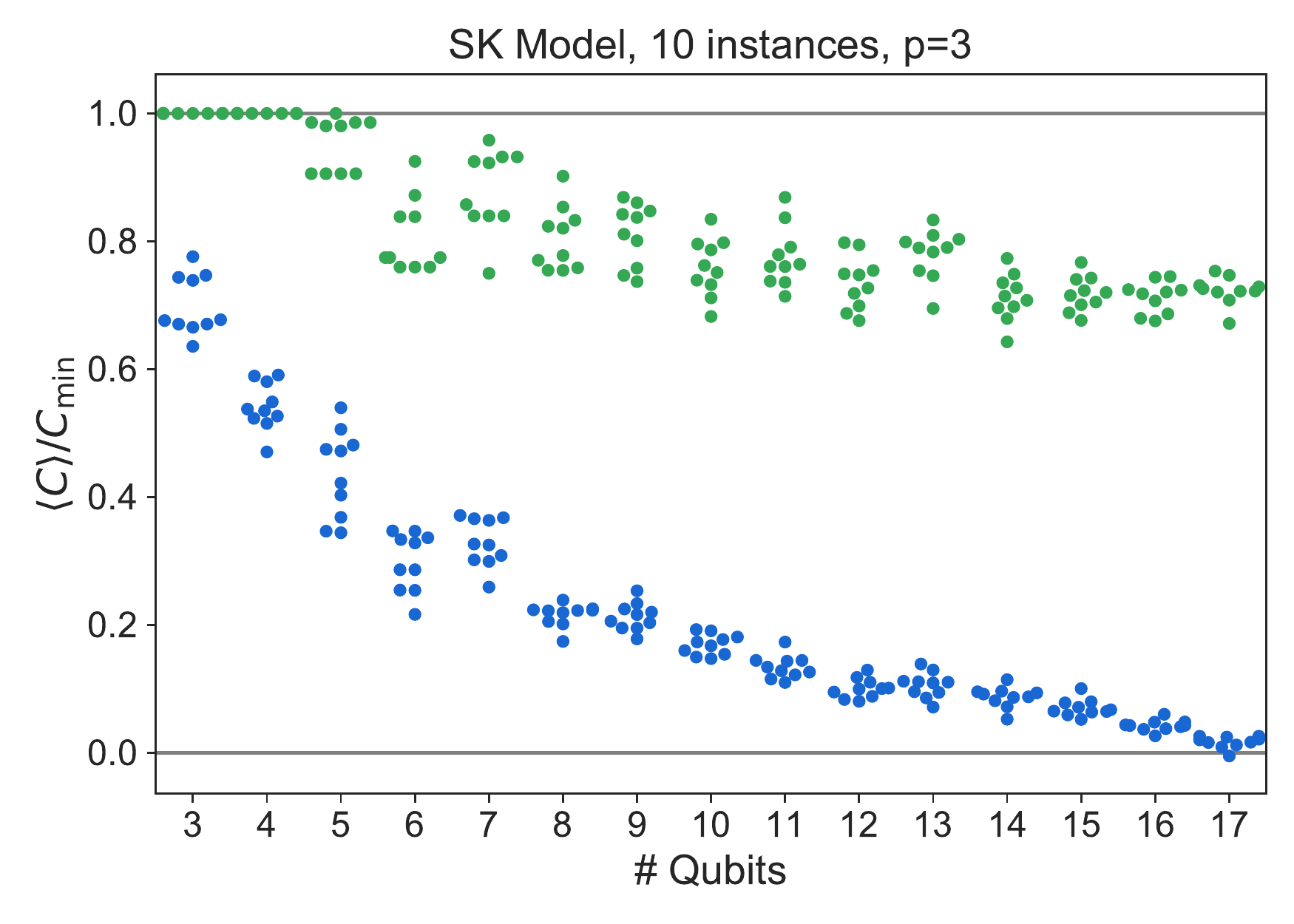}
    \caption{
    Performance of QAOA at $p \in [1, 3]$ and $n \in [3, 17]$ over random SK model instances as described in the main text. Points have been perturbed along the $x$-axis to avoid overlap.
    \textbf{Green:} Noiseless
    \textbf{Blue:} Experimental
    }
    \label{fig:expectation-sk-all}
\end{figure*}

\begin{figure*}[htbp!]
    \centering
    \includegraphics[width=\spfpaoawidth]{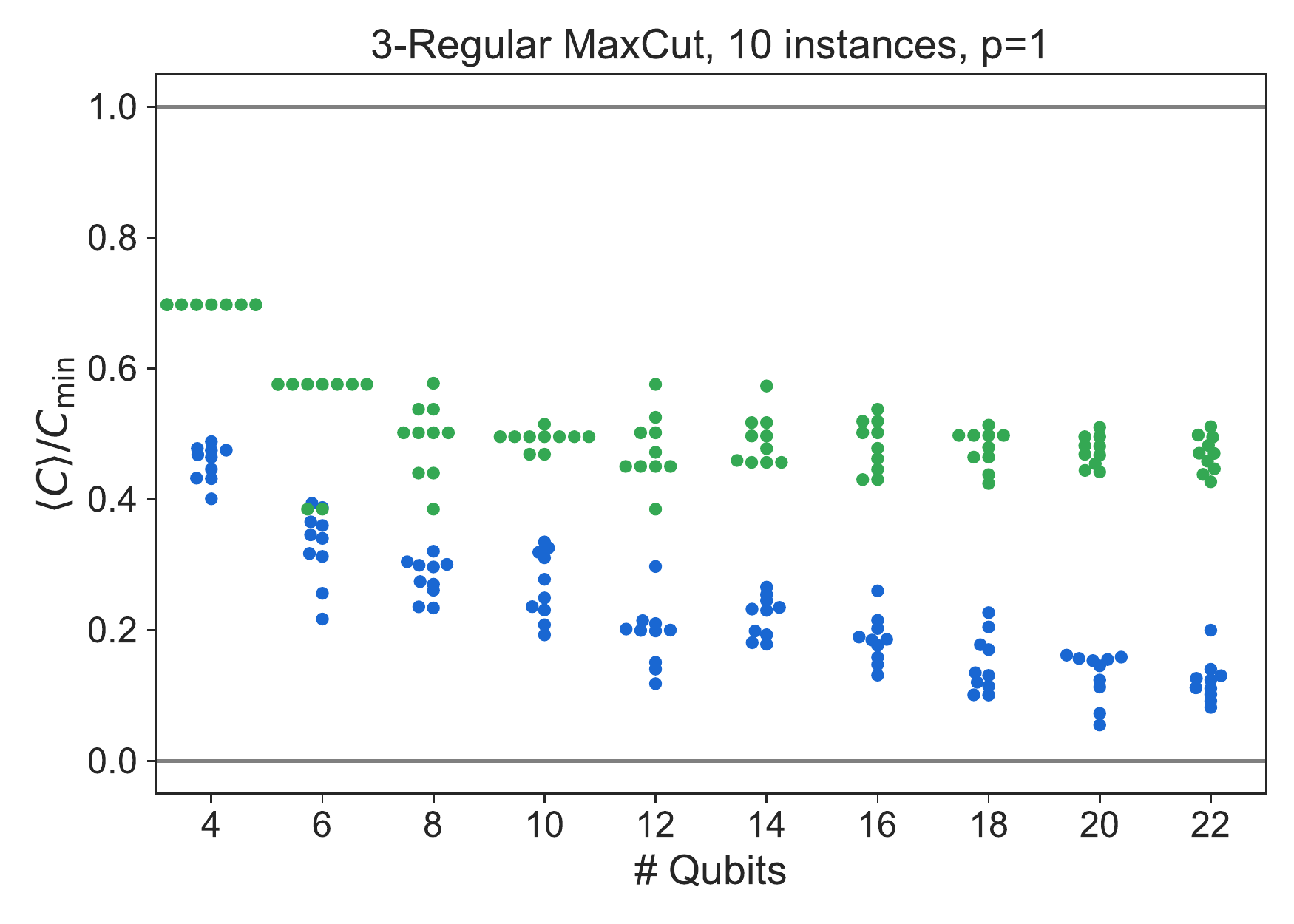}
    \includegraphics[width=\spfpaoawidth]{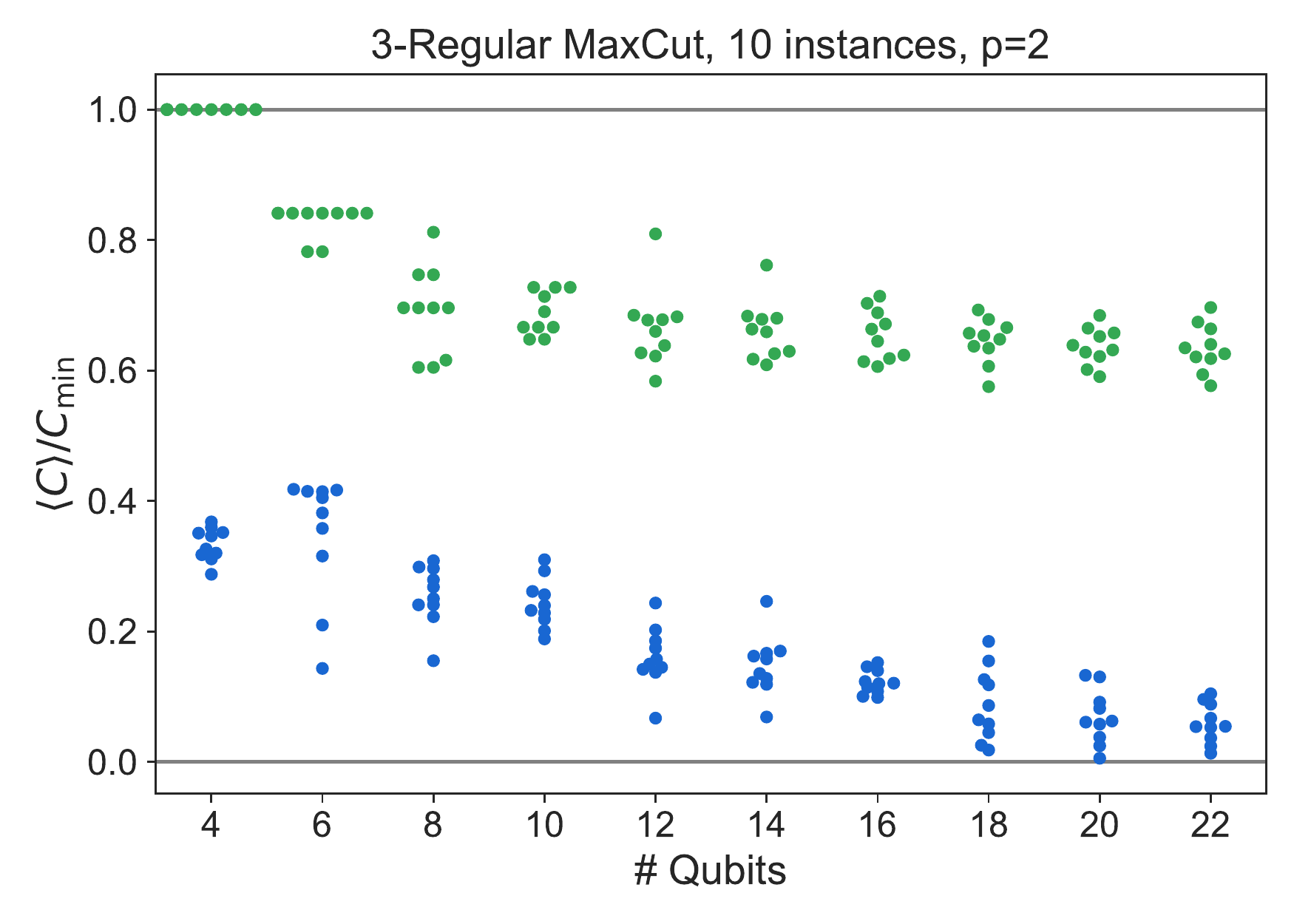}
    \includegraphics[width=\spfpaoawidth]{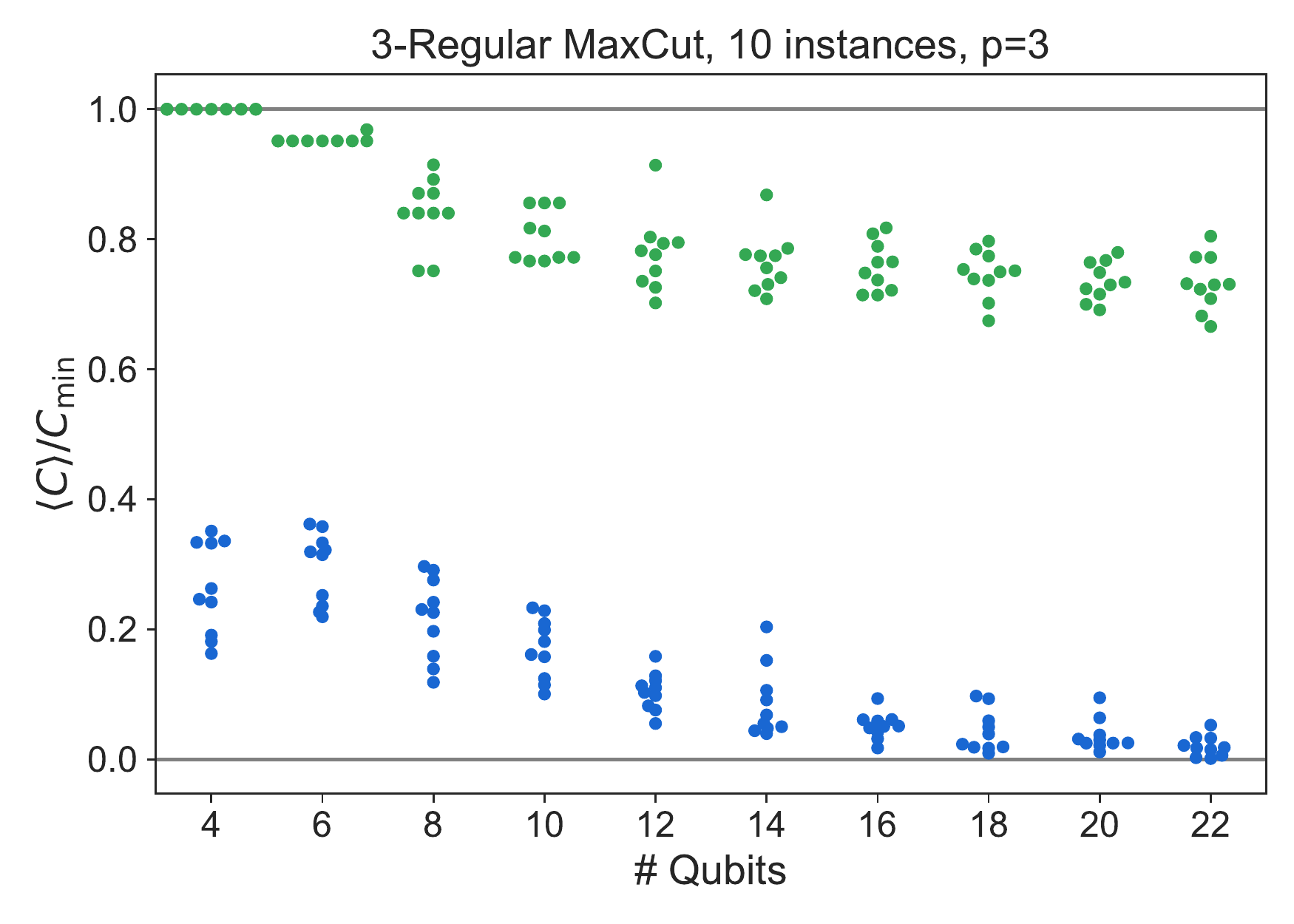}
    \caption{
    Performance of QAOA at $p \in [1, 3]$ and $n \in [4, 22]$ over random 3-regular MaxCut problems as described in the main text. Points have been perturbed along the $x$-axis to avoid overlap.
    $k$-regular graphs must satisfy $n \ge k + 1$ and $nk$ must be even, hence only even $n$ are considered here.
    \textbf{Green:} Noiseless
    \textbf{Blue:} Experimental
    }
    \label{fig:expectation-3-reg-all}
\end{figure*}

\end{document}